\documentclass[10pt]{article}
\usepackage[margin=1in]{geometry}
\usepackage[english]{babel}
\usepackage{amsmath,amssymb,amsthm}

\providecommand{\undertitle}{}
\providecommand{\headeright}{}

\newcounter{chapter}

\RequirePackage{expl3}
\PassOptionsToPackage{dvipsnames}{xcolor}
\usepackage{hyperref}
\usepackage{algorithm}
\usepackage{algpseudocode}  \makeatletter
\renewcommand{\theALG@line}{\thealgorithm.\arabic{ALG@line}} \AtBeginDocument{\def\theHALG@line{\thealgorithm.\arabic{ALG@line}}} \makeatother
\usepackage{mathtools} \usepackage{mathrsfs} \usepackage{enumitem}
\usepackage[font=small]{caption}
\usepackage{float}
\usepackage{tikz}
\usepackage{pgfplots}
\pgfplotsset{compat=1.18}
\usepackage{subcaption}
\usepackage[most]{tcolorbox}
\usetikzlibrary{decorations.pathmorphing} \usetikzlibrary{decorations.pathreplacing} \usetikzlibrary{calc} \usetikzlibrary{fit, backgrounds}
\usetikzlibrary{positioning}
\usetikzlibrary{arrows.meta}
\usepackage{array}
\usepackage{booktabs}
\usepackage{xltabular} \usepackage{cleveref}
\usepackage{comment}
\usepackage{changepage} \usepackage{lipsum} 

\usepackage[acronym]{glossaries}
\glsdisablehyper
\makeglossaries

  \setlist[enumerate]{topsep=0pt, itemsep=0pt}
\setlist[itemize]{topsep=0pt, itemsep=0pt}

\newtcolorbox{helpbox}{
  colback=gray!10,
  colframe=gray!50,
  boxrule=0.5pt,
  arc=4pt,
  left=6pt, right=6pt, top=6pt, bottom=6pt,
  fonttitle=\bfseries,
  coltitle=black,
  title=Help wanted
}

\newtcolorbox{todobox}{
  colback=gray!10,
  colframe=gray!50,
  boxrule=0.5pt,
  arc=4pt,
  left=6pt, right=6pt, top=6pt, bottom=6pt,
  fonttitle=\bfseries,
  coltitle=black,
  title=TODO
}

\newtcolorbox{remarkbox}[1][]{
  colback=yellow!10,
  colframe=yellow!50!black,
  coltitle=black,
  fonttitle=\bfseries,
  title=Remark,
  boxrule=0.8pt,
  arc=3pt,
  left=6pt, right=6pt, top=6pt, bottom=6pt,
  #1
}

\usepackage{aliascnt}
\newtheorem{theorem}{Theorem}[chapter]
\newcommand{\newsharedtheorem}[3]{\newaliascnt{#1}{theorem}\newtheorem{#1}[#1]{#2}\expandafter\def\csname the#1\endcsname{\thetheorem}\expandafter\def\csname theH#1\endcsname{\theHtheorem}\crefname{#1}{\MakeLowercase{#2}}{\MakeLowercase{#3}}\Crefname{#1}{#2}{#3}}
\newsharedtheorem{proposition}{Proposition}{Propositions}
\newsharedtheorem{corollary}{Corollary}{Corollaries}
\newsharedtheorem{axiom}{Axiom}{Axioms}
\newsharedtheorem{claim}{Claim}{Claims}
\newsharedtheorem{assumption}{Assumption}{Assumptions}
\newsharedtheorem{condition}{Condition}{Conditions}
\newsharedtheorem{lemma}{Lemma}{Lemmas}
\newsharedtheorem{example}{Example}{Examples}
\newsharedtheorem{remark}{Remark}{Remarks}
\newsharedtheorem{definition}{Definition}{Definitions}
\newsharedtheorem{problem}{Problem}{Problems}
\newsharedtheorem{conjecture}{Conjecture}{Conjectures}

\newcommand{\Max}{\mathop{\mathrm{Max}}}

\newcommand{\bbN}{\mathbb{N}}

\newcommand{\bbR}{\mathbb{R}}
\newcommand{\bbT}{\mathbb{T}}

\newcommand{\cA}{\mathcal{A}}
\newcommand{\cB}{\mathcal{B}}
\newcommand{\cC}{\mathcal{C}}

\newcommand{\cX}{\mathcal{X}}

\newcommand{\pow}[1]{2^{#1}}
\newcommand{\Dec}{\cX}

\newcommand{\bset}{\cB}

\newcommand{\beh}{b}

\newcommand{\traj}{\tau}                             \newcommand{\trajset}{\cX_{\timeset}}

\newcommand{\timeset}{\bbT}

\newcommand{\agentset}{\mathcal{A}}
\newcommand{\agentpool}{A}
\newcommand{\agent}{a}
\newcommand{\env}{e}

\newcommand{\envset}{\mathcal{E}}
\newcommand{\sceneset}{\mathcal{S}} \newcommand{\scene}{s}              \newcommand{\ruleset}{\mathcal{R}}
\newcommand{\Rule}{R}

\newcommand{\perfset}{\mathcal{Q}}

\newcommand{\resourceset}{\cC}

\newcommand{\resourceorder}{\preceq_{\resourceset}}

\newcommand{\perforder}{\preceq_{\perfset}}
\newcommand{\strictperforder}{\prec_{\perfset}}

\definecolor{posetcolor}{named}{orange}
\newcommand{\posetcolor}{\textcolor{posetcolor}{\(\blacksquare\)}}

\newcommand{\perfposet}{\textcolor{posetcolor}{\perfset}}

\newcommand{\resourceposet}{\textcolor{posetcolor}{\resourceset}}

\newcommand{\fleet}{\mathsf{F}}                        \newcommand{\sysm}{\Psi}                               \newcommand{\metricset}{\mathcal{P}}                   \newcommand{\metricposet}{\textcolor{posetcolor}{\metricset}}
\newcommand{\metricorder}{\preceq_{\metricset}}
\newcommand{\admpools}{\mathbb{A}_{\mathrm{adm}}}      \newcommand{\fleetadm}{\mathbb{F}_{\mathrm{adm}}}      \newcommand{\allformers}{\mathsf{AF}_{\mathrm{all}}}   \newcommand{\deps}{\Delta}                             \newcommand{\bperf}{\pi}

\definecolor{morphismcolor}{named}{blue}

\newcommand{\simmap}{\mathfrak{B}}
\newcommand{\depmap}{\mathfrak{D}}   \newcommand{\deployment}{\delta}   \newcommand{\perf}{\rho}
\newcommand{\Perf}{\mathsf{Perf}}

\newcommand{\Cost}{\mathsf{Cost}}

\newcommand{\safebehset}{\cB_{\mathrm{safe}}}
\newcommand{\af}{\mathrm{af}} \newcommand{\agentformer}{\mathsf{AF}}

\newcommand{\admset}{\cA_{\mathrm{adm}}}
 \newcommand{\lowerset}[1]{\textcolor{posetcolor}{L}#1}

\newcommand{\Reach}{\mathsf{Prof}}   \newcommand{\IReach}{\overline{\Reach}}      \newcommand{\simmapk}[1]{\simmap^{#1}}
\newcommand{\admsetk}[1]{\admset^{#1}}
\newcommand{\Perfk}[1]{\Perf^{#1}}
\newcommand{\afadmk}[1]{\agentformer_{\mathrm{adm}}^{#1}}
\newcommand{\Reachk}[1]{\Reach^{#1}}
\newcommand{\IReachk}[1]{\IReach^{#1}}

\newcommand{\Li}{\mathrm{Li}}

\algnewcommand\algorithmicgiven{\textbf{Given:}}
\algnewcommand\Given{\item[\algorithmicgiven]}

\newglossaryentry{lr}{
    name={legal relation},
    description={...},
}
\newacronym{rdp}{RDP}{rule design problem}
\newacronym{hrdp}{HRDP}{homogeneous rule design problem}
\newacronym{pcp}{PCP}{performance-cost profile}
\newacronym{av}{AV}{Autonomous Vehicle}

\usepackage{chngcntr}
\counterwithin{theorem}{section}

\title{Untitled}
\author{Matteo Penlington}
\date{\today}

\newcommand{\chapter}[1]{\title{#1}\maketitle }

\begin{document}
 \hypersetup{hidelinks}  

\renewcommand{\undertitle}{The Rule Design Problem}
\renewcommand{\headeright}{Working Draft}
\makeatletter
\renewcommand\@fnsymbol[1]{\ifcase#1\or\or 1\or 2\fi}
\makeatother
\author{Matteo Penlington\thanks{This work was supported as a part of NCCR Automation, a National Centre of Competence in Research, funded by the Swiss National Science Foundation (grant number 51NF40\_225155).}\thanks{Institute for Dynamic Systems and Control, ETH Zurich, 8092 Zurich, Switzerland (e-mail: \{mpenlington, lfiaschi, acensi, efrazzoli\}@ethz.ch).} \and Lorenzo Fiaschi\footnotemark[2] \and Raj Kiriti Valcheti\thanks{Department of Electrical and Computer Engineering and Coordinated Science Laboratory at the University of Illinois, Urbana-Champaign, Urbana, IL 61801, USA. (e-mail: {rkv4}@illinois.edu)} \and Andrea Censi\footnotemark[2] \and Emilio Frazzoli\footnotemark[2]}
\date{}

\IfFileExists{stubs.tex}{ }{}

\chapter{Towards a Theory for the Design of Rules in Multi-Agent Systems: the case of Homogeneous Fleets}
\label{ch:rdp_homogeneous}

\begin{center}
    \textbf{Abstract}
\end{center}
\begin{adjustwidth}{2em}{2em}

The regulation of systems composed of many interacting decentralised agents is a long-standing problem.
This problem has typically been resolved by the imposition of rules.
Rules resolve the limitations of sub-optimal agents whilst simultaneously aligning the behaviour of the agents with the demands of society.
Traffic laws, for example, let a driver cross a junction without seeing around the corner, and transport protocols such as TCP let applications share a network without knowing of one another.
The problem is more pressing today, as autonomous systems are deployed in society. 
In this work, we establish the basis of a system-agnostic theory of rule design for multi-agent systems.
We first state the rule design problem (RDP) in its generality. It relates a rule to the set of system performances, defined by the regulator, that the agents deployed under it can attain. 
However, it is not yet possible to solve the RDP in full generality. As such, we introduce the homogeneous rule design problem (HRDP), in which the agents in each environment are identical, and among other interesting properties, show that under worst-case aggregation the two problems have the same solution, bypassing the combinatorial complexity of the RDP. 
To solve the RDP under worst-case aggregation we propose Rule Branch and Bound (\textsc{RuleBnB}), an anytime branch-and-bound algorithm. Finally, we illustrate the value of solving the RDP on an autonomous driving case study.

\end{adjustwidth}

 \section{Introduction}

A fundamental challenge in building and maintaining a functional society lies in the design of commonly agreed structures that regulate the interactions amongst its agents. These structures, referred to as laws, regulations or guidelines, and here as \emph{rules}, serve at least two purposes. First, they compensate for the limitations of decentralised agents, which act on partial information about the system they operate in. Second, they provide the regulator with a means of aligning the behaviour of these agents with the demands of society. Traffic laws illustrate both. Right-of-way rules compensate for the fact that a driver cannot see around a corner nor know the intent of another vehicle, whereas bus lanes reflect the societal demand for public transport that is fast, reliable and cheap, and hence more accessible for all social classes.

The rules in force today have been derived predominantly for human agents, by introspection into our own decision-making and by iteration. The first traffic laws were introduced over 125 years ago~\cite{UK1896LocomotivesAct} and are still regularly amended~\cite{DfT2025HighwayCodeUpdates}. However, today, autonomous agents are increasingly entering decentralised multi-agent systems, both existing ones with existing rules, such as autonomous vehicles on public roads, and new ones that require new rules, such as the transport protocols that govern communication between applications. 
This calls for a scalable and systematic approach to the design of rules.

A classical control perspective on the design of autonomous agents yields decision-making that we understand and behavioural properties that we can guarantee. In this perspective a rule is a constraint imposed on the agent, and the agent selects its behaviour by optimising its objective subject to that constraint. This presupposes that the rule is given. For systems whose interactions we can analyse comprehensively, the rule may be derivable. However, as agents are integrated into ever larger systems of heterogeneous decentralised agents, their interactions can no longer be analysed comprehensively, and the rules they must abide by can neither be readily derived, evaluated nor adjusted, as our case study illustrates (\Cref{sec:case-studies}). This work therefore addresses the question, \emph{given the agents that will operate in a system, which rules should be imposed such that system performance is maximised?}

The question does not admit a straightforward answer. A rule restricts what agents can do, and thereby reduces what they can achieve. However, a rule may also induce coordination. Traffic lanes, for instance, coordinate drivers since each knows that oncoming traffic will not cross into their lane. Moreover, unless some agent design handles everything a system requires, which rule is best depends on the agents that will operate under it.

In this work, we introduce the~\gls{rdp}, a framework for the design of rules in multi-agent systems of decentralised autonomous agents. 
Given a set of available agents, the \gls{rdp} searches for the rule under which the agents that comply with it operate safely and attain the greatest performance. 
The framework treats agents as black boxes and makes no assumption on their structure. The~\gls{rdp} is therefore system- and agent-agnostic, such that from its perspective the design of a transport protocol is no different from the design of the rules of the road.

\subsection{Related Work}
\label{sec:related-work}
The idea of automating the problem of identifying the rules for autonomous systems is not new. 
Several lines of work have formulated particular instances of the~\gls{rdp} within their own formalisms and solved them for specific settings, whereas others have pursued system autonomy through the design of the agents.

\paragraph{Designing rules for explicitly modelled agents.}
A related branch of literature regards the design of social laws and norms for a set of agents whose decision-making is explicitly modelled, as a transition system or as rational preferences~\cite{Shoham1992,Shoham1995SocialLaws}. They use a social law in the same way as we do a rule. 
These works choose the rule so that every agent can still reach its goals~\cite{Shoham1992}, select the least restrictive such rule~\cite{Fitoussi2000SocialLaws,Morales2015Liberal}, optimise the rule against several weighted objectives~\cite{Agotnes2010}, require it to hold when only some of the agents comply~\cite{Agotnes2008Robust} or when the agents are rational~\cite{Bulling2016NormMechanism}, and synthesise or revise it on-line from observed conflicts~\cite{Morales2013,DellAnna2022NormRevision}. 
The~\gls{rdp} is a general abstraction of these in three respects. First, the agents are treated as black boxes, so the framework is agnostic to the system they operate in. Second, no assumption is imposed on the system, and in particular we formulate the problem in a fashion that reflects the objectives of autonomous systems. For example, the least restrictive rule is the objective in~\cite{Fitoussi2000SocialLaws,Morales2015Liberal}, whereas \Cref{thm:nonoptimal} shows that, if the goal is to maximise system performance, without further assumptions, it is not known a priori whether the smallest or the largest rule is best. Third, the system objective is not required to be a scalar. The~\gls{rdp} admits partially ordered performances, characterises a rule by the set of performances attainable under it, which we call its profile, and provides a solver that operates on such sets.

\paragraph{Designing rules for black-box agents.}
Mechanism design~\cite{Sandholm2003} formulates a bilevel optimisation problem that maximises the designer's welfare over the incentives offered to a set of self-interested agents, subject to the agents playing an equilibrium of the game the incentives induce. Mechanism design can be read as a rule design problem in which the incentive scheme plays the role of the rule and the agents' equilibrium behaviour that of the behaviours deployed under it. Various works have likewise learned or optimised a rule for a given set of agents, by imitation and evolutionary reinforcement learning for traffic~\cite{Zhang2024NeuralTrafficRules}, by two-level reinforcement learning for taxation~\cite{Zheng2022AIEconomist}, by simulation-based parameter search for norms~\cite{Montes2022ValueAligned} and by revision from execution traces~\cite{DellAnna2022NormRevision}, and have found that rule-following agents may outperform unconstrained ones. 
These works are potential methods of solving the~\gls{rdp}, and do so for a specific case. In each of them, however, every agent remains in the system whatever the rule, and the rule only shifts the agents' performance.

In the works above, agents always comply with the imposed rules. For social laws, an agent can only execute the actions the rules permit, so it abides by them by construction. For mechanism design, following the mechanism is each agent's best response, so compliance is induced. In practice, however, the rules are inputs to the agents, and when the rules change an agent may no longer be compliant, or no longer safe, and must then be excluded. The~\gls{rdp} treats compliance and safety as properties observed under deployment, and permits only the agents that exhibit them. This distinction matters because the object of interest is the performance the system attains under a rule, not the identity of the agents that attain it, so the set of agents permitted to operate is an outcome of the rule rather than an input to the problem, which is the perspective of a regulator rather than of a designer.

\paragraph{The Agent Design Problem.}
The classical perspective on the design of autonomous systems is to take a given system objective and design the agents so that the objective is met.
Recent work incorporates the rules directly into the policy design of an agent, that is seeking a policy that is optimal under the rules by construction, whether the rules are formalised in temporal logic~\cite{Maierhofer2022,Lin2023MPRobustness}, ordered by priority~\cite{Tumova2013,Censi2019b,Veer2022RecedingVehicles,Lee2025OrderedPreference} or parameterised for a regulator to fix~\cite{Shalev-Shwartz2017}.
However, as discussed in~\cite{penlington2026lexicographicdescentgradientbasedapproach}, rule satisfaction by construction is in general not yet attainable, since each application brings difficulties of its own. For autonomous vehicles one such difficulty is partial observation, as a vehicle cannot see around a corner nor know the intent of another. 
The~\gls{rdp} addresses such intrinsic difficulties of the agent design problem directly through the rules. If the rules are designed so that the agents need not resolve partial observation, as traffic lights do at a junction, then the agents need not be designed to resolve it either. In the agent design problem, the rule is an input to the design and the agent is its output, whereas in the~\gls{rdp} the agents are supplied and the rule is the output. The two problems are thus complementary. 

\subsection{Contributions}
We make four contributions.
\begin{enumerate}
    \item We formalise the~\gls{rdp} in full generality (\Cref{sec:rdp-formal}). This formulation describes rule design for autonomous systems, and presents the objects necessary to pose a well-defined problem.
\item We introduce the~\gls{hrdp}, a special case of the~\gls{rdp} in which all agents in the system are copies of a single agent.
    This will serve as a preliminary step to bypass the intrinsic complexity of~\gls{rdp}.
    \item We then study how the rules of the~\gls{hrdp} shape the performance of the system under worst-case aggregation (\Cref{sec:hrdp}). We find that adding agents to the system can only lower the worst-case attainable performance (\Cref{thm:solo-dominates}), and that, provided the designer has access to every ``best'' agent, the worst-case attainable performance is monotone in the rule (\Cref{thm:optimal-monotone}). The~\gls{hrdp} then has a trivial solution, namely the most permissive rule. However, when the designer lacks some of the ``best'' agents, which as we will see is usually the case, no such monotone relationship exists (\Cref{thm:nonoptimal}). Lastly, we show that under worst-case aggregation a solution to the~\gls{hrdp} is also a solution to the~\gls{rdp} (\Cref{sec:hrdp-bounds-rdp}).
    \item We introduce \textsc{RuleBnB} (\Cref{sec:solving-rdp}), an anytime branch-and-bound \gls{rdp} solver that exploits the ordering of rules by permissiveness, together with the upper bound on the attainable performance from \Cref{sec:hrdp}, to skip evaluating rules that cannot improve on the best performance found so far.
\end{enumerate}
We apply the framework to the regulation of robotaxis (\Cref{sec:case-studies}), where neither the most permissive nor the strictest rule is optimal, an instance of the non-monotone regime of \Cref{thm:nonoptimal}, and show that the solution is interpretable for the regulator and for the agent designer alike.

 \section{The Rule Design Problem, informally}
\label{sec:rdp_informal}
\glsresetall

We now introduce the rule design problem, informally. The aim of this section is to convey the intended semantics of the objects within the rule design problem and how they fit together to form a system-agnostic framework for rule design.

Throughout, we adopt the perspective of a rule designer, by which we mean an entity that has the authority to set the rules that those operating in an environment must abide by, be it a regulator, site operator, or a company.

\subsection{Example 1: Autonomous Vehicles} \label{sec:rdp_inf_ex2}
We introduce the case study used throughout the work. A campus, i.e., a closed site, currently carries no traffic, and several robotaxi suppliers wish to operate within it to transport people between locations. The site's operator has provided us with a brief: the robotaxis must be safe, in that they do not collide with one another, and they must transport people well and at low cost to the operator. Although the operator has no control over what the suppliers offer, it has control over how the robotaxis may behave on its site, and thereby which robotaxis may operate within it. This enables the site operator to hold every robotaxi to the same standard without inspecting the vehicles themselves. Since several robotaxi suppliers may share the site, whether the brief is met depends on how their vehicles behave together. The operator has therefore hired us, \emph{the rule designers}, to find the rules under which the robotaxis meet the brief.

Each robotaxi supplier offers a range of vehicles, much as a car is offered in a basic, a mid-range and a premium trim. The trims differ in their sensors, their computing hardware and their driving software, and hence in price. We call a vehicle an \emph{agent} and the set of vehicles on offer the \emph{catalogue}. Vehicles operate in \emph{environments}, which differ in their scenes, such as the road layout, the time of day and the weather, and in their capacity, the maximum number of vehicles on the roads at once. When vehicles operate in an environment they produce \emph{behaviours}, stochastic joint trajectories.

Rules constrain the behaviours of the vehicles. A speed limit or lane keeping are examples of rules, and each induces the set of behaviours it \emph{permits}. The operator's brief likewise designates as \emph{safe} the behaviours in which no collision occurs. Since we are only interested in vehicles that are safe and execute permitted behaviours, we call a vehicle that is so in every environment \emph{admissible}. Importantly, whether a vehicle is admissible depends on which other vehicles are present, since two vehicles that are each safe on their own may collide once they share the site. Hence, by choosing the rule, we implicitly select which vehicles may operate, and consequently what performance and cost the site attains.

A deployment of vehicles attains a \emph{system performance}, which in this example has two parts, how well people are transported and what it costs the operator. The former is measured on behaviours, by the fraction of requested trips completed or the passengers' waiting time, and the latter on vehicles, by the price of the trim deployed. Each admissible vehicle thus attains a system performance, the least performance it achieves across the environments at its cost, and in general no single vehicle is both the best performing and the cheapest. The \emph{profile} of a rule is the set of system performances attained by its admissible vehicles, together with every system performance that is worse, since attaining one attains those. In the \gls{rdp} we select a rule whose profile is maximal, that is, one contained in the profile of no other candidate.

\begin{problem}[The \glsentrylong{rdp}, informally]
    \label{prob:rdp-informal}
    \begin{center}
    \begin{tabular}{@{}r@{\quad}p{0.78\linewidth}@{}}
        \emph{Given:} & a set of environments; the catalogue of vehicles; the operator's safety requirement, a system performance measure; and a set of candidate rules. \\[0.5ex]
        \emph{Find:} & a candidate rule whose profile is maximal.
    \end{tabular}
    \end{center}
\end{problem}

\begin{figure}[H]
    \centering
    \includegraphics[width=0.95\textwidth]{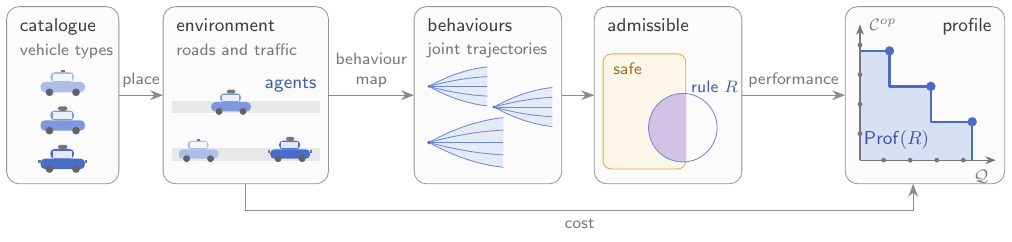}
    \caption{Overview of the components of the informal rule design problem for the robotaxi example. Vehicles (agents) placed in an environment give rise to behaviours, the operator's safety requirement and the rule designate the admissible ones, and performance is measured on behaviours and cost on agents, giving the rule's profile.}
    \label{fig:informal-overview}
\end{figure}

\section{The Rule Design Problem}
\label{sec:rdp-formal}

We now formalise the~\gls{rdp}. To do so, we abstract away from the classical agent design problem by treating agents as black boxes and taking the perspective of a regulator. We build the formalisation to capture how varying a rule changes the system performance that these agents can guarantee.

\subsection{Preliminaries}
\label{sec:prelim}

We use the following notation for order theoretic concepts. A \emph{complete lattice} $\langle P, \leq \rangle$ is a partially ordered set in which every subset $S \subseteq P$ has a meet $\bigwedge S$ and a join $\bigvee S$; in particular $\bigwedge \emptyset = \top$ and $\bigvee \emptyset = \bot$. Throughout, a set written in orange (\posetcolor), such as $\metricposet$, is a complete lattice with the order introduced alongside it; the same set in black, $\metricset$, is its underlying set. The \emph{opposite} $P^{op}$ carries the reversed order. The \emph{downward closure} of $S \subseteq P$ is $\mathop{\downarrow} S \coloneqq \{x \in P \mid \exists\, s \in S,\ x \leq s\}$, a set with $\mathop{\downarrow} S = S$ is a \emph{lower set}, and $\lowerset{P} \subseteq \pow{P}$ denotes the lower sets of $P$. Ordered by inclusion, $\lowerset{P}$ is a complete lattice whose joins are unions and whose meets are intersections. For a poset $\langle P, \leq \rangle$, a set $X$ and a map $f : X \to P$, $\Max_{x \in X} f(x)$ denotes the set of $x \in X$ whose value is maximal, that is, for all $x' \in X$, $f(x) \leq f(x')$ implies $f(x') = f(x)$.

For a set $Z$ and $j \in \bbN$, $Z^{j} \coloneqq \prod_{i=1}^{j} Z$ denotes the $j$-fold Cartesian product.
For ease of exposition, every set carrying a probability measure is equipped with an appropriate $\sigma$-algebra.
Every map between such sets is assumed measurable. For a measurable space $\Omega$, $\mathscr{P}(\Omega)$ denotes the set of probability measures on $\Omega$.

\subsection{Agents, Environments and Behaviours}
\label{sec:rdp-language}

As rule designers, we have no influence over the internal design of the agents within a system, and therefore treat agents as black boxes, with $\agentset$ denoting the set of agents, each of which may be deployed more than once at a time. Agents operate in environments. The \emph{scene} of an environment collects its operating conditions, such as the map geometry, disturbances, weather, and the tasks that the agents may be asked to perform, and $\sceneset$ denotes the set of scenes. An environment has a \emph{capacity}, the maximum number of agents that may be simultaneously operating within it.
\begin{definition}[Environments]
    \label{def:environment}
        An \emph{environment} is a pair $\env \coloneqq (\scene, n)$, where $\scene \in \sceneset$ is a scene and $n \in \bbN_{\geq 1}$ is its capacity. The set of environments is a nonempty set $\envset \subseteq \sceneset \times \bbN_{\geq 1}$ that is downward closed in the capacity, i.e.\ $(\scene, n) \in \envset$ and $k \le n$ imply $(\scene, k) \in \envset$, and has bounded capacity, $N \coloneqq \max\{ n \mid (\scene, n) \in \envset \} < \infty$. We write $[N] \coloneqq \{1, \dots, N\}$.
\end{definition}

Agents operating in an environment produce behaviours. Each agent evolves in a state space over time, and owing to noise, the same agents operating under identical conditions may trace different trajectories. A behaviour abstracts this to the distribution of the joint trajectory.

\begin{definition}[Trajectories]
\label{def:trajectories}
    Let $\timeset$ be a set of times and $\Dec$ the state space of a single agent\footnote{The state space $\Dec$ is common to all agents, as without loss of generality,  distinct state spaces embed into their disjoint union.}. The set of \emph{trajectories} of a single agent is $\trajset \coloneqq \{\traj \mid \traj : \timeset \to \Dec\}$. A \emph{joint trajectory} of $j \in [N]$ agents is a tuple $(\traj^1, \dots, \traj^j) \in \trajset^{\,j}$, where $\traj^i$ is the trajectory of the $i$-th agent.
\end{definition}

\begin{definition}[Behaviour space]
    \label{def:behaviour-space}
    A \emph{behaviour} of $j \in [N]$ agents is a probability measure $\beh \in \bset^{j} \coloneqq \mathscr{P}(\trajset^{\,j})$. The \emph{behaviour space}
    $
        \bset \;\coloneqq\; \bigsqcup_{j = 1}^{N} \bset^{j}
    $
    collects the behaviours of up to $N$ agents.
\end{definition}

In an environment $(\scene, n)$, the agents that may be present are drawn from a set $\agentpool \subseteq \agentset$, the \emph{agent pool} and at most $n$ of them are deployed at once, as a tuple $(a_1, \dots, a_j) \in \agentpool^{j} \subseteq \agentset^{j}$ with $j \le n$. We are interested in all behaviours that could arise in an environment, that is, all behaviours that could arise in its scene under any such tuple. The deployment map gives the behaviours of an arbitrary tuple of agents in a scene, and the behaviour map collects these over all tuples from $\agentpool$ of at most $n$ agents.

\begin{definition}[Deployment map]
    \label{def:deployment-map}
    Let $\agentset^{\le N} \coloneqq \bigcup_{j=1}^{N} \agentset^{j}$. The \emph{deployment map} $\depmap : \agentset^{\le N} \times \sceneset \to \pow{\bset} \setminus \{\emptyset\}$ assigns to a tuple $(a_1, \dots, a_j) \in \agentset^{\leq N}$ and a scene $\scene$ the set $\depmap((a_1, \dots, a_j), \scene) \subseteq \bset^{j}$ of behaviours the agents can produce when deployed in $\scene$.
\end{definition}
\begin{definition}[Behaviour map]
    \label{def:simulation-map-new}
    The \emph{behaviour map} $\simmap : \pow{\agentset} \times \envset \to \pow{\bset}$ is
    \[
        \simmap(\agentpool, (\scene, n)) \;\coloneqq\; \bigcup_{j = 1}^{n}\ \bigcup_{(a_1, \dots, a_j) \in \agentpool^{j}} \depmap\bigl((a_1, \dots, a_j), \scene\bigr).
    \]
\end{definition}

Since the tuples in \Cref{def:simulation-map-new} may repeat agents, a pool accounts for every multiplicity of its members. Furthermore, as $\envset$ is downward closed in the capacity and the union grows with $n$, for every $\agentpool \subseteq \agentset$, $(\scene, n) \in \envset$ and $k \leq n$,
\begin{equation}
    \label{eq:capacity}
    \simmap(\agentpool, (\scene, k)) \;\subseteq\; \simmap(\agentpool, (\scene, n)).
\end{equation}
Allowing more agents in an environment can therefore only add behaviours. This is restrictive only when an environment requires a minimum number of agents to operate at once. Such a lower bound $n_{\min} \le n$ is accommodated by taking environments $(\scene, n_{\min}, n)$ and the union in \Cref{def:simulation-map-new} over $n_{\min} \le j \le n$. The results that follow are unchanged, except \Cref{sec:solo-dominates} and the results that build on it, which require $n_{\min} = 1$.

\subsection{Rules, Safety and Admissibility}
\label{sec:admissibility}

Rules are usually stated as constraints, such as a speed limit or lane keeping. Each such constraint induces the set of behaviours that satisfy it, and under a speed limit every behaviour under which no agent exceeds the speed limit is permitted. Names such as ``speed limit'' or ``lane keeping'' therefore describe sets of permitted behaviours, and in this work a rule is the set it induces.
\begin{definition}[Rules]
\label{def:rule}
    A \emph{rule} is a set $\Rule \subseteq \bset$ of permitted behaviours.
\end{definition}
The rule $\Rule$ is the object we seek as rule designers. As we will see in the following subsections, the rule determines which agents are permitted to operate in an environment, and consequently which system performance is attainable. Although rules such as a speed limit are typically required to hold almost surely, a rule may also be chance-constrained, for instance a speed limit respected with probability at least $0.99$.
\begin{example}[Speed limit]
\label{ex:speed-limit}
    For $j \in [N]$ let $\mathsf{Spd}^{j} \subseteq \trajset^{\,j}$ be the set of joint trajectories in which some agent exceeds the speed $v_{\lim}$. The speed limit is the rule
    $\Rule \coloneqq \bigsqcup_{j = 1}^{N} \bigl\{\, \beh \in \bset^{j} \;\bigm|\; \beh(\mathsf{Spd}^{j}) = 0 \,\bigr\}$,
    and for $\varepsilon \in [0, 1]$ its chance-constrained counterpart replaces $\beh(\mathsf{Spd}^{j}) = 0$ by $\beh(\mathsf{Spd}^{j}) \le \varepsilon$.
\end{example}

In practice several rules are in force at once. The intersection of rules is again a rule, and we take $\Rule$ to be the intersection of all rules in force. For ease of exposition, we refer to $\Rule$ as either ``the rule'' or ``the rules''. As this is an introduction to the~\gls{rdp}, we restrict ourselves to rules as hard constraints and return to prioritised rules, as in rulebooks~\cite{Censi2019b}, in~\Cref{sec:future-work}.

For the theoretical study in~\Cref{sec:prop_ma_systems}, we compare rules by inclusion, and say that $\Rule'$ is \emph{at least as permissive} as $\Rule$ when $\Rule \subseteq \Rule'$.

\subsubsection*{Safety and Admissibility}

We require that the deployed agents are safe and comply with the rules in force, i.e., only execute behaviours that are permitted under $\Rule$. We call a pool of agents that is both safe and compliant \emph{admissible}, and only admissible pools of agents may be deployed. Admissibility is a joint requirement on the pool; it can loosely be read as an assume-guarantee contract~\cite{Benveniste2018ContractsDesign} in which each agent assumes the others are safe and compliant and guarantees the same of itself.

\begin{definition}[Safe behaviours]
\label{def:safe-behaviours}
    The \emph{safe behaviours} are a set $\safebehset \subseteq \bset$.
\end{definition}
Formally, $\safebehset$ has the same structure as a rule, but we keep it separate since safety is fixed by the problem.

\begin{example}[Safety in autonomous driving]
\label{ex:acceptability-AV}
    For $j \in [N]$ let $\mathsf{Col}^{j} \subseteq \trajset^{\,j}$ be the set of joint trajectories in which some agent collides. Collision avoidance is the requirement that collisions have probability zero, $\safebehset \coloneqq \bigsqcup_{j = 1}^{N} \bigl\{\, \beh \in \bset^{j} \;\bigm|\; \beh(\mathsf{Col}^{j}) = 0 \,\bigr\}$.
\end{example}

\begin{definition}[Admissible pool]
\label{def:admissible-pool}
    Let $\env \in \envset$ and $\Rule \subseteq \bset$. A nonempty agent pool $\agentpool \subseteq \agentset$ is \emph{safe} in $\env$ if $\simmap(\agentpool, \env) \subseteq \safebehset$, \emph{compliant} in $\env$ under $\Rule$ if $\simmap(\agentpool, \env) \subseteq \Rule$, and \emph{admissible} if it is both. The admissible pools of $\env$ under $\Rule$ are
    \[
        \admpools(\env, \Rule) \coloneqq \bigl\{\, \agentpool \subseteq \agentset \;\bigm|\; \agentpool \neq \emptyset,\ \simmap(\agentpool, \env) \subseteq \safebehset \cap \Rule \,\bigr\}.
    \]
    
\end{definition}

Since the behaviours of the pair need not be among the behaviours of either alone, two agents each admissible on their own need not be admissible together.

\subsubsection*{Agent Formers and Fleets}

An agent, as introduced in~\Cref{sec:rdp-language}, is a black-box system. The agent deployed may depend on the environment it operates in and on the rules in force. An operator of autonomous vehicles may deploy a vehicle with different sensors, actuators and policy on a closed campus than on a highway, and may retune the policy when the speed limit is lowered. We call the entity that, given an environment and a rule, returns the agent to deploy an \emph{agent former}.

\begin{definition}[Agent former \& catalogue]
    \label{def:agent-former}
        An \emph{agent former} is a function $\af : \envset \times \pow{\bset} \to \agentset$. The \emph{catalogue} $\agentformer$ is a set of agent formers.
\end{definition}

The catalogue collects the formers available to the operator. Two manufacturers with different vehicle ranges and planning stacks, one planning with RRT* and the other with model predictive control, are two formers of the catalogue, as they may return different agents for the same environment and rule.

Several formers may be deployed at once. We call a set of formers deployed together a fleet; under a rule, a fleet deploys in each environment the pool of agents its formers return.

\begin{definition}[Fleet]
    \label{def:fleet}
    A \emph{fleet} is a set $\fleet \subseteq \agentformer$ of agent formers. Under $\Rule \subseteq \bset$, the fleet deploys in $\env \in \envset$ the pool
    \[
        \fleet(\env, \Rule) \;\coloneqq\; \{\, \af(\env, \Rule) \mid \af \in \fleet \,\} \;\subseteq\; \agentset .
    \]
    
\end{definition}

Since only admissible agent pools may be deployed, we also consider only admissible fleets, i.e., the fleets that always deploy an admissible pool.
\begin{definition}[Admissible fleet]
\label{def:admissible-fleet}
    A fleet $\fleet$ is \emph{admissible} under $\Rule \subseteq \bset$ if $\fleet(\env, \Rule) \in \admpools(\env, \Rule)$ for all $\env \in \envset$. Let $\fleetadm(\Rule) \subseteq \pow{\agentformer}$ denote the admissible fleets of the catalogue.
\end{definition}
Since pools are nonempty, $\emptyset \notin \fleetadm(\Rule)$.

\begin{lemma}[Admissibility of fleets is downward closed]
    \label{lem:fleet-admissible-downclosed}
    Let $\Rule \subseteq \bset$. If $\fleet \in \fleetadm(\Rule)$ and $\emptyset \neq \fleet' \subseteq \fleet$, then $\fleet' \in \fleetadm(\Rule)$.
\end{lemma}
\begin{proof}
    Fix $\env \in \envset$. By \Cref{def:fleet}, $\fleet'(\env, \Rule) \subseteq \fleet(\env, \Rule)$. The union in \Cref{def:simulation-map-new} ranges over tuples drawn from the pool, so a smaller pool yields fewer tuples and $\simmap(\fleet'(\env, \Rule), \env) \subseteq \simmap(\fleet(\env, \Rule), \env) \subseteq \safebehset \cap \Rule$, the last inclusion by admissibility of $\fleet$. Since $\fleet' \neq \emptyset$, $\fleet'(\env, \Rule) \neq \emptyset$. Hence $\fleet'(\env, \Rule) \in \admpools(\env, \Rule)$ for every $\env$.
\end{proof}

An \emph{agent deployment} is a family $\deployment = (\agentpool_e)_{e \in \envset} \in (\pow{\agentset})^{\envset}$ of pools, one per environment. To formalise what a catalogue $\agentformer$ can attain, an admissible agent deployment under $\Rule$ picks one admissible pool for each environment.

\begin{definition}[Admissible agent deployment]
    \label{def:admissible-deployment}
    An \emph{admissible agent deployment} under $\Rule$ is a family $\deployment = (\agentpool_e)_{e \in \envset}$ with $\agentpool_e \in \admpools(e, \Rule)$ for every $e \in \envset$. We write $\deps(\Rule)$ for the set of admissible agent deployments under $\Rule$.
\end{definition}
Every admissible fleet $\fleet \in \fleetadm(\Rule)$ induces the admissible agent deployment $\deployment_{\fleet} \coloneqq (\fleet(e, \Rule))_{e \in \envset}$.

\subsection{System Performance and the Profile of a Rule}
\label{sec:rule-reach}

We evaluate a deployment by a single system performance value, which aggregates, in whatever way the designer chooses, the behaviours the deployed agents produce, the agents themselves and the environments they serve. We impose no additional structure on this aggregation beyond its values forming a complete lattice, so that deployments can be compared and aggregated.

\begin{definition}[System performance]
    \label{def:system-metric}
    Let $\metricposet = \langle \metricset, \metricorder \rangle$ be a complete lattice of system performance values, with $m \metricorder m'$ read as $m'$ being at least as desirable as $m$. The \emph{system performance} is a map
    \[
        \sysm : (\pow{\agentset})^{\envset} \times \pow{\bset} \;\longrightarrow\; \metricset
    \]
    that assigns a system performance value to each agent deployment under each rule. For an admissible fleet we write $\sysm(\fleet, \Rule) \coloneqq \sysm(\deployment_{\fleet}, \Rule)$.
    
\end{definition}

\begin{example}
    \label{ex:perf-cost-metric}\label{def:performance-lattice}\label{def:resource-lattice}
    
    One viewpoint is to frame operator requirements as a standard of performance to be achieved at no more than a given cost. Performance is measured on the behaviours the agents produce, and cost is measured on the agents deployed. Let $\perfposet = \langle \perfset, \perforder \rangle$ be a complete lattice of performance values, with $q \perforder q'$ read as $q'$ being at least as desirable as $q$, and $\resourceposet = \langle \resourceset, \resourceorder \rangle$ a complete lattice of cost values, with $c \resourceorder c'$ read as $c$ costing at most as much as $c'$. A \emph{performance map} $\bperf : \bset \times \envset \to \perfset$ assigns a performance value to each behaviour in each environment, and a \emph{cost map} $\Cost : \agentset \to \resourceset$ assigns a cost value to each agent. Let $\metricposet \coloneqq \perfposet \times \resourceposet^{op}$ with the componentwise order, so that $(q, c) \metricorder (q', c')$ if and only if $q \perforder q'$ and $c' \resourceorder c$. For an agent deployment $\deployment = (\agentpool_e)_{e \in \envset}$, the worst case over behaviours, agents and environments gives the system performance 
    \[
        \sysm(\deployment, \Rule) \;\coloneqq\; \Bigl(\, \bigwedge_{e \in \envset}\ \bigwedge_{\beh \in \simmap(\agentpool_e, e)} \bperf(\beh, e),\;\; \bigvee_{e \in \envset}\ \bigvee_{\agent \in \agentpool_e} \Cost(\agent) \Bigr).
    \]
    
\end{example}

A deployment need not be induced by any fleet of the catalogue. We call the catalogue complete when every admissible deployment is weakly dominated by some fleet of the catalogue.

\begin{definition}[Complete catalogue]
    \label{def:complete-catalogue-general}\label{def:complete-catalogue}
    The catalogue $\agentformer$ is \emph{complete} if, for every $\Rule \subseteq \bset$ and every admissible agent deployment $\deployment \in \deps(\Rule)$, there exists an admissible fleet $\fleet \in \fleetadm(\Rule)$ that attains a system performance at least as good, that is, $\sysm(\deployment, \Rule) \metricorder \sysm(\fleet, \Rule)$.
\end{definition}
Every fleet value $\sysm(\fleet, \Rule)$ is attained by the deployment $\deployment_{\fleet}$, so fleet values are among the attainable values $\{\sysm(\deployment, \Rule) \mid \deployment \in \deps(\Rule)\}$. When every attainable value lies below a maximal attainable value, as when $\agentset$ and $\envset$ are finite, the catalogue is complete if and only if every maximal attainable value is attained by an admissible fleet. Since values in $\metricposet$ may be incomparable, there is in general no greatest attainable value, only maximal ones.

We now introduce the main object by which we evaluate how ``good'' a rule is. Fix a rule $\Rule$. Every admissible fleet $\fleet$ attains the system performance $\sysm(\fleet, \Rule)$. If this value is $m$, then an operator who deploys $\fleet$ under $\Rule$ is guaranteed a performance of at least $m$, and so it also guarantees every value $m' \metricorder m$. We call the set of all values some admissible fleet guarantees under $\Rule$ the profile of $\Rule$, the downward closure of the attained values.

\begin{definition}[Profile of a rule]
    \label{def:profile}
    The \emph{profile} is the set-valued map
    $
        \Reach : \pow{\bset} \longrightarrow \lowerset{\metricset}
    $
    defined by
    \[
        \Reach(\Rule) \;\coloneqq\; \mathop{\downarrow} \Bigl\{\, \sysm(\fleet, \Rule) \;\Bigm|\; \fleet \in \fleetadm(\Rule) \,\Bigr\}.
    \]
    
\end{definition}

\begin{figure}[H]
    \centering
    \begin{subfigure}[b]{0.48\textwidth}
        \centering
        \includegraphics{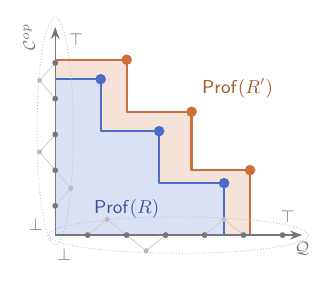}
        \caption{$\Reach(\Rule) \subseteq \Reach(\Rule')$.}
        \label{fig:formal-reach-nested}
    \end{subfigure}
    \hfill
    \begin{subfigure}[b]{0.48\textwidth}
        \centering
        \includegraphics{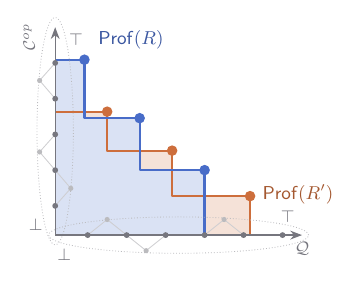}
        \caption{Profiles are not comparable.}
        \label{fig:formal-reach-incomparable}
    \end{subfigure}
    \caption{Comparing rules by their profiles, drawn for the system performance of \Cref{ex:perf-cost-metric}. (\subref{fig:formal-reach-nested}) The profile of $\Rule'$ contains the profile of $\Rule$, so every pair guaranteed under $\Rule$ is guaranteed under $\Rule'$. (\subref{fig:formal-reach-incomparable}) Each profile contains a pair the other lacks. $\Rule$ guarantees a cheaper fleet and $\Rule'$ a better-performing one, and the two rules are incomparable.}
    \label{fig:formal-reach-comparison}
\end{figure}

Our goal as rule designers is to pick a rule $\Rule$. To do so, we need to compare $\Rule$ with another rule $\Rule'$ and understand which is better. Each rule has a profile, $\Reach(\Rule)$ and $\Reach(\Rule')$, and since both are lower sets of $\metricset$, we can compare them by inclusion. If $\Reach(\Rule) \subseteq \Reach(\Rule')$ (\Cref{fig:formal-reach-nested}), then every value guaranteed under $\Rule$ is also guaranteed under $\Rule'$, so $\Rule'$ is at least as good as $\Rule$, and strictly better if the inclusion is strict; as rule designers we should then pick $\Rule'$. If instead neither profile contains the other (\Cref{fig:formal-reach-incomparable}), $\Reach(\Rule) \not\subseteq \Reach(\Rule')$ and $\Reach(\Rule') \not\subseteq \Reach(\Rule)$, then neither rule is dominant and it is up to the rule designer and the operators to jointly decide which rule is best.

We can thus introduce the Rule Design Problem (RDP), in which, as rule designers, we seek to identify a rule whose profile is maximal.
\begin{problem}[\Glsentrylong{rdp}]
    Given $\Reach : \pow{\bset} \to \lowerset{\metricset}$, find a rule whose profile is maximal under inclusion:
    \begin{equation}
        \Max_{\Rule \subseteq \bset} \quad \Reach(\Rule).
        \tag{RDP}\label{prob:RDP}
    \end{equation}
\end{problem}
A maximal profile exists whenever the rules considered form a finite set, as in \Cref{sec:solving-rdp}, and need not exist otherwise.

Often in practice, as rule designers, we are provided with a certain system performance to be attained. If this is the case, we are then interested in the following bounded rule design problem.

\begin{problem}[Bounded \glsentrylong{rdp}]
    Given further a requirement $L^* \in \lowerset{\metricset}$, find a rule whose profile is maximal under inclusion among those that serve it:
    \begin{equation}
        \begin{aligned}
            \Max_{\Rule \subseteq \bset} &\quad \Reach(\Rule) \\
            \text{s.t.} &\quad L^* \subseteq \Reach(\Rule).
        \end{aligned}
        \tag{BRDP}\label{prob:BRDP}
    \end{equation}
    
\end{problem}
A required value $m^* \in \metricset$ is the requirement $L^* = \mathop{\downarrow}\{m^*\}$, for which the constraint reads $m^* \in \Reach(\Rule)$.

\subsection{Future-Proofing the RDP and its Difficulties}
\label{sec:licensing}

In principle the catalogue would contain every former that could ever be deployed. However, only the formers of today are known, so we solve \eqref{prob:RDP} with the current catalogue $\agentformer$, and as technology develops the question becomes which new formers may be added to a deployed fleet. Let $\allformers \supseteq \agentformer$ denote the set of all formers, known or not, with fleets over $\allformers$ and their admissibility defined as in \Cref{sec:admissibility} with $\allformers$ in place of $\agentformer$.

One option is the closure operator of \cite{li2026compatibilityfutureprooftestingmultiagent}, stated here with ``compatible'' read as ``admissible''.
\begin{definition}[Closure~\cite{li2026compatibilityfutureprooftestingmultiagent}]
    \label{def:closure}
    Let $\fleet \subseteq \allformers$ be admissible under $\Rule \subseteq \bset$. The \emph{closure} $\Li_{\Rule}(\fleet)$ of $\fleet$ is the intersection of all admissible fleets that contain $\fleet$ and are maximal under inclusion.
\end{definition}
By \cite[Prop.~1 and Property~3]{li2026compatibilityfutureprooftestingmultiagent}, such maximal fleets exist, so $\Li_{\Rule}$ is well defined, and it is extensive, monotone and idempotent; in particular $\fleet \subseteq \Li_{\Rule}(\fleet)$, and $\Li_{\Rule}(\fleet)$ is admissible by \Cref{lem:fleet-admissible-downclosed}. It is not to be confused with the downward closure of \Cref{sec:prelim}.

By \Cref{lem:fleet-admissible-downclosed}, any former in $\Li_{\Rule}(\fleet)$ may be added to $\fleet$, alone or together with any others in $\Li_{\Rule}(\fleet)$, without affecting admissibility. Performance may change, since the guarantee becomes $\sysm(\fleet \cup G, \Rule)$ for the formers $G$ added.

Thus, one solves \eqref{prob:RDP} with the current catalogue, commissions, for a solution $\Rule^{*}$ of \eqref{prob:RDP}, an admissible fleet $\fleet$ whose performance $\sysm(\fleet, \Rule^{*})$ is the chosen guarantee, and admits a new former if and only if it belongs to $\Li_{\Rule^{*}}(\fleet)$. Computing $\Li_{\Rule}(\fleet)$ is future work.

Solving \eqref{prob:RDP} in this future-proof form is at present not possible for two reasons. First, the closure cannot be computed, since it is taken over all formers and rests on the joint behaviours of agents that do not yet exist. 
Second, the system performance of a fleet containing agents that do not yet exist cannot be evaluated or bounded for the same reason. 

We can, however, compute the profile over the formers of the current catalogue.
Identifying the ``correct'' concept for fleet compatibility needs greater investigation. Thus, as a starting case, in the next section we study the RDP for the case where homogeneous fleets are deployed.

\section{The Homogeneous Rule Design Problem}
\label{sec:hrdp}

\subsection{Problem Formulation}
\label{sec:hrdp-formulation}

In the homogeneous setting a single former is deployed at a time. The deployed pool is then a singleton $\{\agent\}$, every tuple in \Cref{def:simulation-map-new} is of the form $(\agent, \dots, \agent)$, and admissibility reduces to the agent being safe and compliant when deployed alongside itself. We abbreviate $\simmap(\agent, \env) \coloneqq \simmap(\{\agent\}, \env)$, and likewise for the other objects defined on a pool. \Cref{def:admissible-pool} then reads as follows.

\begin{definition}[Admissible agent]
\label{def:admissible-agent}
    Let $\env \in \envset$ and $\Rule \subseteq \bset$. The \emph{safe} agents of $\env$ and the \emph{compliant} agents of $\env$ under $\Rule$ are
    \[
        \agentset_{\mathrm{safe}}(\env) \coloneqq \bigl\{\, \agent \in \agentset \;\bigm|\; \simmap(\agent, \env) \subseteq \safebehset \,\bigr\},
        \qquad
        \agentset_{\mathrm{comp}}(\env, \Rule) \coloneqq \bigl\{\, \agent \in \agentset \;\bigm|\; \simmap(\agent, \env) \subseteq \Rule \,\bigr\}.
    \]
    The \emph{admissible} agents of $\env$ under $\Rule$ are
    \[
        \admset(\env, \Rule) \coloneqq \agentset_{\mathrm{safe}}(\env) \cap \agentset_{\mathrm{comp}}(\env, \Rule) = \bigl\{\, \agent \in \agentset \;\bigm|\; \simmap(\agent, \env) \subseteq \safebehset \cap \Rule \,\bigr\}.
    \]
\end{definition}

\begin{remark}
    Deciding whether a black-box agent is always safe or compliant is itself difficult and an active area of research~\cite{Mathiesen2023NeuralBarrier,Campi2018WaitAndJudge,Badings2022ScenarioMDP,Salamati2024DataDrivenBarrier}. Working towards a unified theory of the design of autonomous systems, we assume that the behaviour set $\simmap(\agent, \env)$ is available, from which admissibility can be readily determined.
\end{remark}
Since only admissible agents may be deployed, we also consider only admissible agent formers, i.e., the agent formers that always output an admissible agent.
\begin{definition}[Admissible agent former]
\label{def:admissible-agent-former}
    An agent former $\af$ is \emph{admissible} under $\Rule \subseteq \bset$ if $\af(\env, \Rule) \in \admset(\env, \Rule)$ for all $\env \in \envset$, and let $\agentformer_{\mathrm{adm}}(\Rule) \subseteq \agentformer$ denote the admissible formers of the catalogue.
\end{definition}
Which formers the catalogue contains is specific to the problem instantiation. In some instances the catalogue matches every admissible deployment (\Cref{def:complete-catalogue-general}); this is not always true. We will study the impact that this has on the solutions to the~\gls{rdp} in~\Cref{sec:prop_ma_systems}.

The \emph{homogeneous profile} is the profile over singleton fleets,
\begin{equation}
    \label{eq:hrdp-profile}
    \Reach_{\mathrm{H}}(\Rule) \;\coloneqq\; \mathop{\downarrow} \bigl\{\, \sysm(\{\af\}, \Rule) \;\bigm|\; \af \in \agentformer_{\mathrm{adm}}(\Rule) \,\bigr\},
\end{equation}
and $\Reach_{\mathrm{H}}(\Rule) \subseteq \Reach(\Rule)$ since a singleton is a fleet.

\begin{problem}[Homogeneous \glsentrylong{rdp}]
    Given $\sysm$ and the catalogue $\agentformer$, find a rule whose profile is maximal under inclusion \eqref{eq:hrdp-profile}:
    \begin{equation}
        \Max_{\Rule \subseteq \bset} \quad \Reach_{\mathrm{H}}(\Rule).
        \tag{HRDP}\label{prob:HRDP}
    \end{equation}
    The bounded form adds the constraint $L^* \subseteq \Reach_{\mathrm{H}}(\Rule)$ as in \eqref{prob:BRDP}.
    
\end{problem}

\subsection{The Properties of Rules under Worst-Case Aggregation}
\label{sec:prop_ma_systems}
In this section we study how the capacity and the rule shape the profile. Since autonomous systems are usually designed against the worst case, we study the case in which system performance is aggregated by its worst case.

\begin{definition}[Worst-case aggregation of system performance]
    \label{def:worst-case}\label{def:per-env-performance}
    A \emph{performance} is a map $\perf : \agentset \times \bset \times \envset \to \metricset$, with $\perf(\agent, \beh, \env)$ the performance attributed to agent $\agent$ under behaviour $\beh$ in environment $\env$. The \emph{per-environment performance} of a pool $\agentpool \subseteq \agentset$ in $\env$ and the system performance are
    \[
        \sysm_{\env}(\agentpool) \;\coloneqq\; \bigwedge_{\agent \in \agentpool}\ \bigwedge_{\beh \in \simmap(\agentpool, \env)} \perf(\agent, \beh, \env),
        \qquad
        \sysm\bigl((\agentpool_e)_{e \in \envset}, \Rule\bigr) \;\coloneqq\; \bigwedge_{e \in \envset} \sysm_{e}(\agentpool_e).
    \]
    As for the behaviour map we abbreviate $\sysm_{\env}(\agent) \coloneqq \sysm_{\env}(\{\agent\})$.
    
\end{definition}

\begin{remark}[Choice of aggregation]
    \label{rem:aggregation}
    The system performance aggregates at two levels, the performance of the behaviours an agent produces in an environment, and the performance of the agents a former deploys across environments. Both aggregators are a modelling choice of the designer. Common classes are the worst case, the best case and the expected case. Which properties of rules hold depends on the choice of aggregation. In this work we adopt the worst case since when designing autonomous systems one typically requires that, once deployed, the agents are guaranteed to meet a pre-defined performance standard.
\end{remark}

We first show that under worst-case aggregation the profile coincides with the homogeneous profile (\Cref{sec:hrdp-bounds-rdp}), so that the results that follow apply to the RDP and the HRDP alike. We then show that the profile is antitone in the capacity, the number of agents that may operate at once (\Cref{sec:solo-dominates}). We next bound what any catalogue can attain by an ideal profile that depends on the rule, and use it to show that under a complete catalogue the profile is monotone in the permissiveness of the rule, so that the most permissive rule solves \eqref{prob:HRDP} (\Cref{sec:pcp-monotone}). Catalogues need not be complete, however, and we show by example that a stricter rule can then strictly enlarge the profile (\Cref{sec:pcp-nonmonotone}).

\subsubsection*{Preliminaries}
Recall that an agent is admissible under a rule $\Rule$ if it is safe and compliant while operating alongside itself (\Cref{def:admissible-agent}). Since the behaviours an agent produces are not linked to the rule, enlarging the rule can only add admissible agents.
\begin{lemma}[Admissibility is monotone in the rule]
    \label{lem:adm-monotone}
    For every $\env \in \envset$ and $\Rule, \Rule' \subseteq \bset$,
    \[
        \Rule \subseteq \Rule' \;\implies\; \admset(\env, \Rule) \subseteq \admset(\env, \Rule').
    \]
\end{lemma}
\begin{proof}
    By \Cref{def:admissible-agent}, $\agent \in \admset(\env, \Rule)$ if and only if $\simmap(\agent, \env) \subseteq \safebehset \cap \Rule$. If $\Rule \subseteq \Rule'$ then $\safebehset \cap \Rule \subseteq \safebehset \cap \Rule'$, and the claim follows by transitivity of $\subseteq$.
\end{proof}

\subsubsection{The HRDP solves the RDP under worst-case aggregation}
\label{sec:hrdp-bounds-rdp}

Under worst-case aggregation, adding a former to a fleet cannot raise the system performance. Thus, the system performance of a fleet is bounded by above by that attained with a homogeneous fleet. Conversely, a single former is itself a fleet, so any system performance attained by a homogeneous fleet is attained in \eqref{prob:RDP} as well. Hence, the~\eqref{prob:HRDP} solves~\eqref{prob:RDP} under worst-case aggregation; we now prove this formally.

\begin{lemma}[Performance is antitone in the pool]
    \label{lem:perf-pool}
    For every $\agentpool \subseteq \agentpool' \subseteq \agentset$ and $\env \in \envset$, $\simmap(\agentpool, \env) \subseteq \simmap(\agentpool', \env)$ and $\sysm_{\env}(\agentpool') \metricorder \sysm_{\env}(\agentpool)$.
    
\end{lemma}
\begin{proof}
    Let $\env = (\scene, n)$. For every $j \le n$ and $(a_1, \dots, a_j) \in \agentpool^{j}$ we have $(a_1, \dots, a_j) \in \agentpool'^{\,j}$, hence $\depmap((a_1, \dots, a_j), \scene) \subseteq \simmap(\agentpool', \env)$ by \Cref{def:simulation-map-new}, and taking the union over $j$ and over the tuples gives $\simmap(\agentpool, \env) \subseteq \simmap(\agentpool', \env)$. Let $S \coloneqq \{\perf(\agent, \beh, \env) \mid \agent \in \agentpool,\ \beh \in \simmap(\agentpool, \env)\}$ and $S' \coloneqq \{\perf(\agent, \beh, \env) \mid \agent \in \agentpool',\ \beh \in \simmap(\agentpool', \env)\}$, so that $\sysm_{\env}(\agentpool) = \bigwedge S$ and $\sysm_{\env}(\agentpool') = \bigwedge S'$ by \Cref{def:worst-case}. Since $\agentpool \subseteq \agentpool'$ and $\simmap(\agentpool, \env) \subseteq \simmap(\agentpool', \env)$, $S \subseteq S'$. Hence $\bigwedge S'$ is a lower bound of $S$, and as $\bigwedge S$ is the greatest lower bound of $S$, $\bigwedge S' \metricorder \bigwedge S$.
    
\end{proof}

\begin{lemma}[Homogeneous fleets upper-bound fleet performance]
    \label{lem:fleet-member}
    Let $\fleet \in \fleetadm(\Rule)$ and $\af \in \fleet$. Then $\af \in \agentformer_{\mathrm{adm}}(\Rule)$ and $\sysm(\fleet, \Rule) \metricorder \sysm(\{\af\}, \Rule)$.
\end{lemma}
\begin{proof}
    Fix $\env$. By \Cref{lem:perf-pool}, $\simmap(\{\af(\env, \Rule)\}, \env) \subseteq \simmap(\fleet(\env, \Rule), \env) \subseteq \safebehset \cap \Rule$, so $\af(\env, \Rule) \in \admset(\env, \Rule)$, and $\sysm_{\env}(\fleet(\env, \Rule)) \metricorder \sysm_{\env}(\af(\env, \Rule))$. Taking the meet over $\env$ gives the claim.
\end{proof}

\begin{theorem}[The HRDP solves the RDP under worst-case aggregation]
    \label{thm:hrdp-bounds-rdp}
    Under \Cref{def:worst-case}, for every catalogue $\agentformer$ and rule $\Rule \subseteq \bset$,
    \[
        \Reach(\Rule) \;=\; \Reach_{\mathrm{H}}(\Rule).
    \]
    
\end{theorem}
\begin{proof}
    $\Reach_{\mathrm{H}}(\Rule) \subseteq \Reach(\Rule)$ since every admissible singleton is an admissible fleet. Conversely, $\Reach_{\mathrm{H}}(\Rule)$ is a lower set, and for every $\fleet \in \fleetadm(\Rule)$ and any $\af \in \fleet$, \Cref{lem:fleet-member} gives $\sysm(\fleet, \Rule) \metricorder \sysm(\{\af\}, \Rule) \in \Reach_{\mathrm{H}}(\Rule)$.
    
\end{proof}

Since the profiles coincide for every rule, \eqref{prob:HRDP} and \eqref{prob:RDP} have the same solutions.

\begin{corollary}[Licensing is bounded by the homogeneous profile]
    \label{cor:licensing}
    Under \Cref{def:worst-case}, with $\agentformer_{\mathrm{adm}}$ and $\Reach_{\mathrm{H}}$ taken over $\allformers$, for every rule $\Rule$, admissible fleet $\fleet$ and $\af \in \Li_{\Rule}(\fleet)$,
    \[
        \sysm\bigl(\Li_{\Rule}(\fleet), \Rule\bigr) \;\metricorder\; \sysm(\fleet, \Rule)
        \qquad\text{and}\qquad
        \sysm\bigl(\Li_{\Rule}(\fleet), \Rule\bigr) \;\metricorder\; \sysm(\{\af\}, \Rule) \;\in\; \Reach_{\mathrm{H}}(\Rule).
    \]
\end{corollary}
\begin{proof}
    $\Li_{\Rule}(\fleet)$ is an admissible fleet containing $\fleet$ (\Cref{def:closure}). The first inequality is \Cref{lem:perf-pool} applied to $\fleet(\env, \Rule) \subseteq \Li_{\Rule}(\fleet)(\env, \Rule)$ in every environment, taking the meet over $\env$; the second is \Cref{lem:fleet-member}, whose proof does not depend on the catalogue, applied to $\af \in \Li_{\Rule}(\fleet)$.
\end{proof}

Three remarks on~\Cref{thm:hrdp-bounds-rdp}.
First, under worst-case aggregation a fleet is scored by its weakest member and its worst joint behaviour, so adding formers can only add terms to the meet, and licensing lowers the guarantee further (\Cref{cor:licensing}).
Second, the ideal profile of \Cref{sec:pcp-monotone} bounds both the HRDP and the RDP, and it does not depend on the catalogue. If a requirement lies outside the ideal profile of every rule, then no catalogue under any rule can meet it, and the problem itself must be revisited, through its tasks, its environments or its safety requirement.
Third, of the two inclusions that make up the equality $\Reach(\Rule) = \Reach_{\mathrm{H}}(\Rule)$, only $\Reach(\Rule) \subseteq \Reach_{\mathrm{H}}(\Rule)$ rests on worst-case aggregation, through \Cref{lem:perf-pool}. The inclusion $\Reach_{\mathrm{H}}(\Rule) \subseteq \Reach(\Rule)$ holds for any aggregation, since singletons are fleets. Arguably, expectation is a more natural aggregator in certain cases, particularly those involving agent cooperation\footnote{One advantage of expectation is that a single very poor behaviour occurring with very low probability has little influence if all the other behaviours are similar.}, and under an expectation a heterogeneous fleet may strictly enlarge the profile. Whether the HRDP still solves the RDP then depends on the underlying distributions, which we leave for future study.

\subsubsection{Single-agent deployments weakly dominate multi-agent deployments}
\label{sec:solo-dominates}

We now show that the profile is antitone in the capacity, so that single-agent deployments weakly dominate multi-agent deployments. For an operator this means that, unless the task itself requires several agents to operate at once, restricting each environment to a single agent never shrinks the set of guaranteed values, and may enlarge it.

Throughout this section, for $k \in [N]$ let $\simmapk{k}(\agentpool, (\scene, n)) \coloneqq \simmap\bigl(\agentpool, (\scene, \min\{k, n\})\bigr)$ be the behaviour map in which every environment is run with at most $k$ agents, so that $k$ bounds the capacity of every environment; we call $k$ the \emph{capacity bound}. Quantities built from $\simmapk{k}$ in place of $\simmap$, and otherwise unchanged, carry the superscript $k$, e.g.\ $\admsetk{k}$, $\sysm^{k}_{\env}$, $\afadmk{k}$ and $\Reachk{k}$. \Cref{lem:perf-pool,lem:fleet-member,thm:hrdp-bounds-rdp} hold verbatim for the superscript-$k$ quantities, since $\simmapk{k}$ is $\simmap$ at a smaller capacity, so $\Reachk{k} = \Reach^{k}_{\mathrm{H}}$.

\begin{lemma}[Admissibility is antitone in the capacity]
    \label{lem:adm-capacity}
    For every $\env \in \envset$, $\Rule \subseteq \bset$ and $k, k' \in [N]$,
    \[
        k \le k' \;\implies\; \admsetk{k'}(\env, \Rule) \subseteq \admsetk{k}(\env, \Rule).
    \]
\end{lemma}
\begin{proof}
    Let $\agent \in \admsetk{k'}(\env, \Rule)$ denote an admissible agent. By \Cref{def:admissible-agent}, agent $\agent$ is admissible if and only if $\simmapk{k'}(\{\agent\}, \env) \subseteq \safebehset \cap \Rule$. By \eqref{eq:capacity}, $k \le k'$ gives $\simmapk{k}(\{\agent\}, \env) \subseteq \simmapk{k'}(\{\agent\}, \env)$, so $\simmapk{k}(\{\agent\}, \env) \subseteq \safebehset \cap \Rule$, i.e.\ $\agent \in \admsetk{k}(\env, \Rule)$.
\end{proof}

\begin{lemma}[Performance is antitone in the capacity]
    \label{lem:perf-capacity}
    For every $\agentpool \subseteq \agentset$, $\env \in \envset$ and $k, k' \in [N]$,
    \[
        k \le k' \;\implies\; \sysm^{k'}_{\env}(\agentpool) \metricorder \sysm^{k}_{\env}(\agentpool).
    \]
\end{lemma}
\begin{proof}
    Let $k \le k'$. By \eqref{eq:capacity}, $\simmapk{k}(\agentpool, \env) \subseteq \simmapk{k'}(\agentpool, \env)$. By \Cref{def:worst-case}, $\sysm^{k'}_{\env}(\agentpool) \metricorder \perf(\agent, \beh', \env)$ for every $\agent \in \agentpool$ and $\beh' \in \simmapk{k'}(\agentpool, \env)$, in particular for every $\beh' \in \simmapk{k}(\agentpool, \env)$. Thus $\sysm^{k'}_{\env}(\agentpool)$ is a lower bound of $\{\perf(\agent, \beh', \env) \mid \agent \in \agentpool,\ \beh' \in \simmapk{k}(\agentpool, \env)\}$, and so lies below its greatest lower bound, $\sysm^{k}_{\env}(\agentpool)$.
\end{proof}

\begin{theorem}[The profile is antitone in the capacity]
    \label{thm:solo-dominates}
    For every catalogue $\agentformer$, $\Rule \subseteq \bset$ and $k, k' \in [N]$,
    \[
        k \le k' \;\implies\; \Reachk{k'}(\Rule) \subseteq \Reachk{k}(\Rule).
    \]
\end{theorem}
\begin{proof}
    Let $\af \in \afadmk{k'}(\Rule)$. By \Cref{lem:adm-capacity}, $\af(\env, \Rule) \in \admsetk{k'}(\env, \Rule) \subseteq \admsetk{k}(\env, \Rule)$ for every $\env$, so $\af \in \afadmk{k}(\Rule)$. Let $m \coloneqq \bigwedge_{\env} \sysm^{k}_{\env}(\af(\env, \Rule))$ and $m' \coloneqq \bigwedge_{\env} \sysm^{k'}_{\env}(\af(\env, \Rule))$, so that $\af$ attains $m$ in $\Reachk{k}(\Rule)$ and $m'$ in $\Reachk{k'}(\Rule)$. By \Cref{lem:perf-capacity}, $m' \metricorder \sysm^{k'}_{\env}(\af(\env, \Rule)) \metricorder \sysm^{k}_{\env}(\af(\env, \Rule))$ for every $\env$, so $m'$ is a lower bound of $\{\sysm^{k}_{\env}(\af(\env, \Rule)) \mid \env \in \envset\}$ and $m' \metricorder m$. Hence $m' \metricorder m \in \Reachk{k}(\Rule)$. So $\Reachk{k}(\Rule)$ is a lower set containing every value attained by a former in $\afadmk{k'}(\Rule)$, and therefore contains their downward closure, $\Reachk{k'}(\Rule) \subseteq \Reachk{k}(\Rule)$.
\end{proof}

Applying~\Cref{thm:solo-dominates} along $1 \le 2 \le \cdots \le N$ results in
\[
    \Reach(\Rule) = \Reachk{N}(\Rule) \;\subseteq\; \Reachk{N-1}(\Rule) \;\subseteq\; \cdots \;\subseteq\; \Reachk{1}(\Rule),
\]
for every catalogue and rule. This shows that, not only does the single-agent profile weakly dominate every other, but the profile can only decrease as the capacity grows, as each additional agent an environment may hold can remove guaranteed values.

This has a direct implication for operators. Since the profile can only grow as the capacity decreases, it is beneficial to design scenes so that as few agents as possible need to operate within them simultaneously, for instance by scheduling so that a single agent serves a scene at a time.

\begin{remark}[Dependence on worst-case aggregation]
    \label{rem:cyclists}
    \Cref{thm:solo-dominates} holds for any aggregator over behaviours that is antitone under inclusion, and may fail otherwise, notably when coordination improves performance. One example of this is cycling, where cyclists may ride alone or together, but typically ride as a group since drafting reduces aerodynamic drag. Under our model, $\simmapk{k'}$ contains both the single- and the multi-agent behaviours, so the worst case is set by riding alone, whereas an expectation, potentially the more natural aggregator here, may increase with the number of cyclists.
\end{remark}

\subsubsection{Profile monotonicity with complete catalogues}
\label{sec:pcp-monotone}

To show that under a complete catalogue the~profile is monotone in the permissiveness of the rule, we introduce the \emph{ideal}~profile, an upper bound on the~profile that depends on the rule, and show that complete catalogues attain it. As a consequence, under a complete catalogue the most permissive rule $\Rule = \bset$ solves \eqref{prob:HRDP}.

The ideal~profile is obtained by disregarding the catalogue and allowing any admissible agent to be deployed in each environment. A value is then guaranteed in an environment if some admissible agent of that environment attains a value at least as good, and it is guaranteed under the rule if this holds in every environment. The ideal~profile is the intersection of these per-environment guarantees.
\begin{definition}[Ideal profile]
    \label{def:ideal-profile}
    The \emph{ideal profile} of a rule $\Rule \subseteq \bset$ is
    \begin{equation}
        \label{eq:majorant}
        \IReach(\Rule) \;\coloneqq\; \bigcap_{\env \in \envset}\ \mathop{\downarrow}\Bigl\{\, \sysm_{\env}(\agent) \;\Bigm|\; \agent \in \admset(\env, \Rule) \,\Bigr\}
        \;\in\; \lowerset{\metricset}.
    \end{equation}
\end{definition}

The ideal~profile has three properties of interest here.
First, it contains the profile of every catalogue, $\Reach(\Rule) \subseteq \IReach(\Rule)$, since a former deploys admissible agents.
Second, it is monotone in the rule, $\Rule \subseteq \Rule' \implies \IReach(\Rule) \subseteq \IReach(\Rule')$, since relaxing the rule adds admissible agents (\Cref{lem:adm-monotone}) and hence new, potentially improved values.
Third, the ideal~profile is attained by complete catalogues. Every value in the ideal profile is dominated by the value of some admissible agent deployment, which under a complete catalogue is in turn dominated by an admissible fleet, so the value lies in the profile; conversely, a catalogue that attains the ideal profile under every rule matches every deployment.
\begin{proposition}[Properties of the ideal profile]
    \label{prop:ideal-profile}
    For every catalogue $\agentformer$ and all $\Rule, \Rule' \subseteq \bset$:
    \begin{enumerate}
        \item $\Reach(\Rule) \subseteq \IReach(\Rule)$;
        \item $\Rule \subseteq \Rule' \implies \IReach(\Rule) \subseteq \IReach(\Rule')$;
        \item $\agentformer$ is complete (\Cref{def:complete-catalogue-general}) if and only if $\IReach(\Rule) = \Reach(\Rule)$ for every $\Rule \subseteq \bset$.
    \end{enumerate}
\end{proposition}
\begin{proof}
    \emph{(i)} For $\env \in \envset$ write $L_\env \coloneqq \mathop{\downarrow}\{\sysm_{\env}(\agent) \mid \agent \in \admset(\env, \Rule)\}$ so that $\IReach(\Rule) = \bigcap_{\env} L_\env$. Since $\Reach(\Rule)$ is the downward closure of the values attained by the formers in $\agentformer_{\mathrm{adm}}(\Rule)$ (\Cref{thm:hrdp-bounds-rdp}) and $\IReach(\Rule)$ is a lower set, it suffices to show that each such value lies in $\IReach(\Rule)$. Let $\af \in \agentformer_{\mathrm{adm}}(\Rule)$, set $\agent_\env \coloneqq \af(\env, \Rule)$, and let $m_{\af} \coloneqq \bigwedge_{\env \in \envset} \sysm_{\env}(\agent_\env)$ be the value it attains. Fix $\env \in \envset$. By \Cref{def:admissible-agent-former}, $\agent_\env \in \admset(\env, \Rule)$, so $\sysm_{\env}(\agent_\env) \in L_\env$. Moreover $m_{\af} \metricorder \sysm_{\env}(\agent_\env)$. As $L_\env$ is a lower set, $m_{\af} \in L_\env$. Since $\env$ was arbitrary, $m_{\af} \in \bigcap_{\env} L_\env = \IReach(\Rule)$.

    \emph{(ii)} By \Cref{lem:adm-monotone}, $\admset(\env, \Rule) \subseteq \admset(\env, \Rule')$ for every $\env$. Each lower set in \eqref{eq:majorant} grows, and so does their intersection.

    \emph{(iii)} ($\Rightarrow$) Suppose $\agentformer$ is complete. The inclusion $\Reach(\Rule) \subseteq \IReach(\Rule)$ is (i). For the reverse inclusion $\IReach(\Rule) \subseteq \Reach(\Rule)$, let $m \in \IReach(\Rule)$. For each $\env$ there is $\agent_\env \in \admset(\env, \Rule)$ with $m \metricorder \sysm_{\env}(\agent_\env)$. Then $\deployment \coloneqq (\{\agent_\env\})_\env$ is an admissible agent deployment under $\Rule$, and since $m$ is a lower bound of $\{\sysm_{\env}(\agent_\env) \mid \env \in \envset\}$, $m \metricorder \bigwedge_{\env} \sysm_{\env}(\agent_\env) = \sysm(\deployment, \Rule)$. By \Cref{def:complete-catalogue-general} there is $\fleet \in \fleetadm(\Rule)$ with $\sysm(\deployment, \Rule) \metricorder \sysm(\fleet, \Rule)$, so $m \metricorder \sysm(\fleet, \Rule)$ and $m \in \Reach(\Rule)$ by \Cref{def:profile}.

    ($\Leftarrow$) Assume $\IReach(\Rule) = \Reach(\Rule)$ for every $\Rule$, and let $\deployment = (\agentpool_\env)_\env$ be an admissible agent deployment under $\Rule$. Pick $a_\env \in \agentpool_\env$ for each $\env$; by \Cref{lem:perf-pool}, $a_\env \in \admset(\env, \Rule)$ and $\sysm(\deployment, \Rule) \metricorder m \coloneqq \bigwedge_{\env} \sysm_{\env}(a_\env)$. For each $\env$, the agent $a_\env$ generates the value $\sysm_{\env}(a_\env)$ of the $\env$-th lower set in \eqref{eq:majorant}, and $m \metricorder \sysm_{\env}(a_\env)$. Hence $m \in \IReach(\Rule) = \Reach(\Rule)$, so by \Cref{def:profile} there is $\fleet \in \fleetadm(\Rule)$ with $\sysm(\deployment, \Rule) \metricorder m \metricorder \sysm(\fleet, \Rule)$, which is the condition of \Cref{def:complete-catalogue-general}, so $\agentformer$ is complete.
\end{proof}

With a complete catalogue, relaxing the rule can only add admissible agents, and every previously admissible deployment remains matched by a former, so the profile cannot decrease.
\begin{theorem}[The profile is monotone in the rule under a complete catalogue]
    \label{thm:optimal-monotone}
    Let the catalogue $\agentformer$ be complete (\Cref{def:complete-catalogue-general}). Then for all $\Rule, \Rule' \subseteq \bset$,
    \[
        \Rule \subseteq \Rule' \;\implies\; \Reach(\Rule) \subseteq \Reach(\Rule').
    \]
\end{theorem}

\begin{proof}
    Let $\Rule \subseteq \Rule'$. By \Cref{prop:ideal-profile}(iii) the profile of a complete catalogue is the same as the ideal profile, and by \Cref{prop:ideal-profile}(ii) the ideal profile is monotone in the rule, so
    $
        \Reach(\Rule) = \IReach(\Rule) \subseteq \IReach(\Rule') = \Reach(\Rule').
    $
\end{proof}

Under a complete catalogue, permitting everything yields the largest~profile under inclusion, so $\Rule = \bset$ solves \eqref{prob:HRDP}. This is because an agent admissible under a stricter rule remains admissible under the relaxed rule (\Cref{lem:adm-monotone}), and a complete catalogue has a former that matches the deployment of such agents.

\subsubsection{The profile is not monotone under incomplete catalogues}
\label{sec:pcp-nonmonotone}
Typically, catalogues are not complete. When the catalogue is incomplete, a former tuned to perform well alone may, under a more permissive rule, switch to an agent that is better alone but unsafe alongside itself. The former is then inadmissible under the relaxed rule, and the profile shrinks.

\begin{theorem}[Incomplete catalogues can break monotonicity]
    \label{thm:nonoptimal}
    There exist an instance, an incomplete catalogue $\agentformer$ and rules $\Rule \subset \Rule' \subseteq \bset$ such that
    \[
        \Reach(\Rule') \;\subset\; \Reach(\Rule).
    \]
\end{theorem}
\begin{proof}
    Take the performance of \Cref{ex:perf-cost-metric}, $\metricset = \perfset \times \resourceset^{op}$ with $\perf(\agent, \beh, \env) = (\bperf(\beh, \env), \Cost(\agent))$, and write $\Perfk{1}(\agent, \env)$ for the first component of $\sysm^{1}_{\env}(\agent)$.
    Let $\sceneset = \{\scene\}$, $N = 2$ and $\envset = \{(\scene,1), (\scene,2)\}$. Let $\agentset = \{a, b, d\}$ with $\Cost(a) = \Cost(b) = c$ and $\Cost(d) = c_d$, and let $\beta_a, \beta_{aa}, \beta_b, \beta_{bb}, \beta_d, \beta_{dd} \in \bset$ be six behaviours with $\beta_{bb} \notin \safebehset$ and the other five in $\safebehset$. Let $\beta_x \in \bset^{1}$ and $\beta_{xx} \in \bset^{2}$ with
    \[
        \depmap((x), \scene) = \{\beta_x\}, \qquad \depmap((x, x), \scene) = \{\beta_{xx}\}, \qquad x \in \{a, b, d\},
    \]
    and $\depmap$ nonempty but otherwise arbitrary on the mixed tuples, which do not enter \eqref{eq:hrdp-profile}, so $\simmap(x, (\scene,1)) = \{\beta_x\}$ and $\simmap(x, (\scene,2)) = \{\beta_x, \beta_{xx}\}$. Let $\bperf(\beta_x, \cdot) = \bperf(\beta_{xx}, \cdot) = q_x$ with $q_d \strictperforder q_a \strictperforder q_b$ in $\perfposet$, so that $b$ performs best alone but is unsafe when multiple instances of itself are present. Let
    \[
        \Rule \coloneqq \{\beta_a, \beta_{aa}, \beta_d, \beta_{dd}\}, \qquad \Rule' \coloneqq \Rule \cup \{\beta_b, \beta_{bb}\}, \qquad \Rule \subset \Rule'.
    \]
    By \Cref{def:admissible-agent}, $\admset(\env, \Rule) = \{a, d\}$ for both $\env$, $\admset((\scene,1), \Rule') = \{a, b, d\}$ and $\admset((\scene,2), \Rule') = \{a, d\}$.

    The catalogue is $\agentformer = \{\af_1, \af_d\}$, where $\af_d(\env, \cdot) \coloneqq d$ is constant and $\af_1$ is the \emph{solo-greedy} former, which deploys the agent of best solo performance $\Perfk{1}(\cdot, \env)$ among $\admsetk{1}(\env, \Rule'')$, and $a$ if that set is empty; this is well defined for every $\Rule''$ since $q_d \strictperforder q_a \strictperforder q_b$ are distinct and form a chain. Since $\Perfk{1}(x, \env) = q_x$ and $\admsetk{1}(\env, \Rule) = \{a, d\}$, $\admsetk{1}(\env, \Rule') = \{a, b, d\}$ for both $\env$, we get $\af_1(\env, \Rule) = a$ and $\af_1(\env, \Rule') = b$.

    Under $\Rule$ both formers are admissible and, by \eqref{eq:hrdp-profile} and \Cref{thm:hrdp-bounds-rdp},
    \[
        \Reach(\Rule) = \mathop{\downarrow}\{(q_a, c), (q_d, c_d)\}.
    \]
    Under $\Rule'$, $\af_1((\scene,2), \Rule') = b \notin \admset((\scene,2), \Rule')$, so $\agentformer_{\mathrm{adm}}(\Rule') = \{\af_d\}$ and
    \[
        \Reach(\Rule') = \mathop{\downarrow}\{(q_d, c_d)\}.
    \]
    As $(q_a, c) \in \Reach(\Rule) \setminus \Reach(\Rule')$ (since $(q_a, c) \in \Reach(\Rule') = \mathop{\downarrow}\{(q_d, c_d)\}$ would require $q_a \perforder q_d$, which together with $q_d \strictperforder q_a$ contradicts antisymmetry) and $\Reach(\Rule') \subseteq \Reach(\Rule)$, we have $\Reach(\Rule') \subset \Reach(\Rule)$, and $\agentformer$ is incomplete by \Cref{thm:optimal-monotone}.
\end{proof}

Since catalogues are typically incomplete, this non-monotone regime is one that as rule designers we must work within. This is also what makes the problem difficult to solve in general. The solution to the~\gls{rdp} depends on the catalogue, the environments, and how the agents behave within them. In the following section we use the bound of \Cref{prop:ideal-profile}(i) to skip rules whose bound cannot improve on the best~profile found so far.
 \section{Solving the RDP under Worst-Case Aggregation}
\label{sec:solving-rdp}

We now introduce \textsc{RuleBnB}, an anytime branch-and-bound solver for the RDP under worst-case aggregation for a finite set of candidate rules. \textsc{RuleBnB} seeks to minimise the number of evaluations of $\Reach(\Rule)$, and does so in three steps. 
First, it computes an upper bound $\IReachk{1}(\Rule)$ on $\Reach(\Rule)$ for every candidate rule, which costs little because it requires the system performance of each agent at capacity one only, and this once for all rules. 
Second, it picks a rule, evaluates its profile, and removes from consideration every rule whose upper bound the evaluated profile dominates, so that it never evaluates those rules (\Cref{sec:bnb}). 
Third, since $\Reach(\Rule)$ is itself expensive to compute, \textsc{RuleBnB} accepts a user-defined approximation of it. It enumerates the evaluation of the agent formers, and as soon as the possible gain that a former can still add towards the upper bound falls below a user-specified threshold, it abandons that former and moves to the next (\Cref{sec:efficient-pcp}).

\begin{assumption}[Finite candidate set]
    \label{ass:finite}
    Let $\langle \ruleset, \subseteq \rangle$ be a finite set of candidate rules $\ruleset \subseteq \pow{\bset}$ ordered by inclusion. The sets $\agentset$, $\agentformer$ and $\envset$ are finite, and the admissible formers $\agentformer_{\mathrm{adm}}(\Rule)$ of every $\Rule \in \ruleset$ are given, since operators certify the formers they offer.
\end{assumption}
Notably, although \textsc{RuleBnB} can in principle integrate the certification of admissible formers, elaborating on this is beyond the scope of this work. For simplicity, the assumption instead delegates the certification of admissibility to another entity, which provides us with $\agentformer_{\mathrm{adm}}(\Rule)$.

Throughout, let $\IReach$ denote $\IReachk{1}$, and we refer to $\simmapk{1}(\{\agent\}, \env)$ as the \emph{single-agent behaviours}. By \Cref{thm:solo-dominates} and \Cref{prop:ideal-profile}, $\Reach(\Rule) \subseteq \IReach(\Rule)$ for every rule, and $\IReach$ is monotone in the rule.

\subsection{Pruning candidate rules}
\label{sec:bnb}

\textsc{RuleBnB} prunes a candidate $\Rule$ whenever its upper bound $\IReach(\Rule)$ lies inside the profile of a rule already evaluated, or does not attain a required set of system performances $L^* \in \lowerset{\metricset}$, where $L^*$ is the requirement of~\eqref{prob:BRDP} and $L^* = \{\bot_{\metricset}\}$ for~\eqref{prob:RDP}.
Let $V \subseteq \ruleset$ be the evaluated rules, each $\Rule' \in V$ with a lower set $\widehat{\Reach}(\Rule') \subseteq \Reach(\Rule')$. For a candidate $\Rule \in \ruleset$,
\begin{equation}
    \textsc{Prune}(\Rule, V) \;\coloneqq\; \bigl[\, L^* \not\subseteq \IReach(\Rule) \,\bigr] \;\lor\; \bigl[\, \exists\, \Rule' \in V \text{ such that } \IReach(\Rule) \subseteq \widehat{\Reach}(\Rule') \,\bigr]. \label{eq:prune}
\end{equation}
We call $\Rule$ \emph{feasible} if $L^* \subseteq \Reach(\Rule)$ and write $\ruleset^* \subseteq \ruleset$ for the feasible candidates. For~\eqref{prob:RDP} every rule with a non-empty profile is feasible, so the first clause prunes only rules with an empty profile (if every candidate has an empty profile, every candidate solves \eqref{prob:RDP} and $\ruleset^* = \emptyset$).
We now show that pruning is safe and down-closed. Safety means that no pruned rule attains a performance that an evaluated rule does not, and down-closedness means that pruning a rule prunes every stricter rule with it.

\begin{lemma}[Pruning is safe and down-closed]
    \label{lem:safe-pruning}
    Let $\Rule \in \ruleset$ and let $V \subseteq \ruleset$ with $\widehat{\Reach}(\Rule') \subseteq \Reach(\Rule')$ for every $\Rule' \in V$.
    \begin{enumerate}
        \item If $\IReach(\Rule) \subseteq \widehat{\Reach}(\Rule')$ for some $\Rule' \in V$, then $\Reach(\Rule) \subseteq \Reach(\Rule')$.
        \item If $L^* \not\subseteq \IReach(\Rule)$, then $\Rule$ is not feasible.
        \item $\textsc{Prune}(\Rule, V)$ implies $\textsc{Prune}(\Rule'', V)$ for every $\Rule'' \in \mathop{\downarrow}\Rule$, the down-set taken in $\langle \ruleset, \subseteq \rangle$.
    \end{enumerate}
\end{lemma}
\begin{proof}
    \emph{(i)} and \emph{(ii)} follow from $\Reach(\Rule) \subseteq \IReach(\Rule)$ (\Cref{thm:solo-dominates,prop:ideal-profile}), together with $\widehat{\Reach}(\Rule') \subseteq \Reach(\Rule')$ for~(i).
    \emph{(iii)} For $\Rule'' \in \mathop{\downarrow}\Rule$, \Cref{prop:ideal-profile}(ii) gives $\IReach(\Rule'') \subseteq \IReach(\Rule)$, so each clause of \eqref{eq:prune} passes from $\Rule$ to $\Rule''$.
\end{proof}

\subsection{Evaluating the profile efficiently}
\label{sec:efficient-pcp}

We now show how to evaluate $\Reach(\Rule)$ by a user-defined approximation. 
By \Cref{thm:hrdp-bounds-rdp} it suffices to evaluate fleets that consist of a single former, and by \Cref{lem:perf-capacity} the performance of a former is bounded from above by the performance of the agents it returns when each is deployed alone. 
\begin{definition}[True performance and single-agent bound of a former]
    \label{def:former-pair}
    Let $\Rule \subseteq \bset$, $\af \in \agentformer_{\mathrm{adm}}(\Rule)$ and $\agent_\env \coloneqq \af(\env, \Rule)$ for $\env \in \envset$. The \emph{true performance} $\psi_{\af} \in \metricposet$ and the \emph{single-agent bound} $\bar\psi_{\af} \in \metricposet$ of $\af$ under $\Rule$ are
    \[
        \psi_{\af} \;\coloneqq\; \bigwedge_{\env \in \envset} \sysm_{\env}(\agent_\env),
        \qquad
        \bar\psi_{\af} \;\coloneqq\; \bigwedge_{\env \in \envset} \sysm^{1}_{\env}(\agent_\env).
    \]
\end{definition}
The evaluation of a profile may terminate after any former, and the evaluation of a former after any environment. 
Let $X$ be the formers evaluated so far and $\envset'$ the environments in which the current former has been evaluated. 
\begin{definition}[Realised profile and running aggregate]
    \label{def:running-aggregate}
    Let $\Rule \subseteq \bset$, $X \subseteq \agentformer_{\mathrm{adm}}(\Rule)$, $\af \in \agentformer_{\mathrm{adm}}(\Rule)$, $\agent_\env \coloneqq \af(\env, \Rule)$ and $\envset' \subseteq \envset$. The \emph{realised profile} with $X$ and the \emph{running aggregate} of $\af$ on $\envset'$ are
    \[
        \widehat{\Reach}_X(\Rule) \;\coloneqq\; \mathop{\downarrow}\{\, \psi_{\af} \mid \af \in X \,\},
        \qquad
        \psi_{\af}(\envset') \;\coloneqq\; \bigwedge_{\env \in \envset'} \sysm_{\env}(\agent_\env).
    \]
\end{definition}

\begin{figure}[!tb]
    \centering
    \begin{subfigure}[b]{0.24\textwidth}
        \centering
        \includegraphics[width=\linewidth]{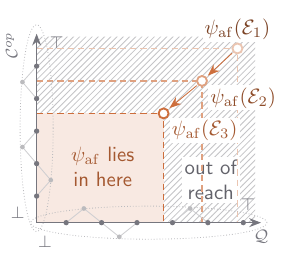}
        \caption{Running aggregate.}
        \label{fig:running-aggregate}
    \end{subfigure}
    \hfill
    \begin{subfigure}[b]{0.24\textwidth}
        \centering
        \includegraphics[width=\linewidth]{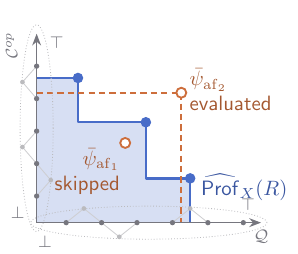}
        \caption{Single-agent bound.}
        \label{fig:majorant-removal}
    \end{subfigure}
    \hfill
    \begin{subfigure}[b]{0.24\textwidth}
        \centering
        \includegraphics[width=\linewidth]{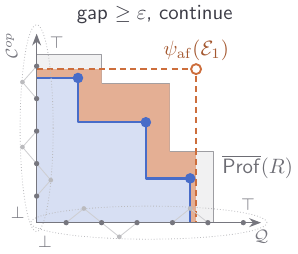}
        \caption{Large gap, continue.}
        \label{fig:gap-continue}
    \end{subfigure}
    \hfill
    \begin{subfigure}[b]{0.24\textwidth}
        \centering
        \includegraphics[width=\linewidth]{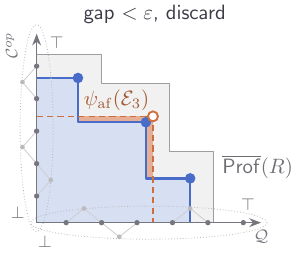}
        \caption{Small gap, discard.}
        \label{fig:gap-discard}
    \end{subfigure}
    \caption{All panels are drawn in the lattice $\metricset = \perfset \times \resourceset^{op}$ of \Cref{ex:perf-cost-metric}. (\subref{fig:running-aggregate}) The running aggregate of a former moves from $\psi_{\af}(\envset_1)$ to $\psi_{\af}(\envset_2)$ and $\psi_{\af}(\envset_3)$ as environments are evaluated, $\envset_1 \subset \envset_2 \subset \envset_3$, lowering the bound. The true performance $\psi_{\af}$ lies in $\mathop{\downarrow} \psi_{\af}(\envset_3)$, and the hatched region is out of reach. (\subref{fig:majorant-removal}) Blue is the profile realised so far. The single-agent bound $\bar\psi_{\af_1}$ lies inside it, so $\af_1$ is skipped, whereas $\bar\psi_{\af_2}$ lies outside it, so $\af_2$ has to be evaluated. (\subref{fig:gap-continue}, \subref{fig:gap-discard}) The gap, shaded orange, is the part of $\mathop{\downarrow} \psi_{\af}(\envset') \cap \IReach(\Rule)$ not covered by the realised profile. After few environments it extends far beyond the realised profile, and after further environments a thin sliver remains, and once its measure is below $\varepsilon$ the former is discarded.}
    \label{fig:former-tests}
\end{figure}

The realised profile grows with $X$, equals $\Reach(\Rule)$ at $X = \agentformer_{\mathrm{adm}}(\Rule)$, and is the $\widehat{\Reach}(\Rule)$ of \Cref{sec:bnb}. 
The running aggregate is non-increasing in the environments evaluated and reaches $\psi_{\af}$ at $\envset' = \envset$ (\Cref{fig:running-aggregate}). 
The single-agent bound and the running aggregate are upper bounds on the true performance.

\begin{lemma}[Bounds on the true performance of a former]
    \label{lem:former-bounds}
    Let $\Rule \subseteq \bset$ and $\af \in \agentformer_{\mathrm{adm}}(\Rule)$. Then $\psi_{\af} \metricorder \bar\psi_{\af}$ and $\psi_{\af} \in \mathop{\downarrow} \psi_{\af}(\envset') \cap \IReach(\Rule)$ for every $\envset' \subseteq \envset$.
\end{lemma}
\begin{proof}
    Let $\agent_\env \coloneqq \af(\env, \Rule)$. By \Cref{lem:perf-capacity}, $\sysm_{\env}(\agent_\env) \metricorder \sysm^{1}_{\env}(\agent_\env)$ for every $\env$, so $\bigwedge_{\env} \sysm_{\env}(\agent_\env)$ is a lower bound of the single-agent performances and lies below their meet, whence $\psi_{\af} \metricorder \bar\psi_{\af}$. For $\envset' \subseteq \envset$, the meet over $\envset$ lies below the meet over $\envset'$, whence $\psi_{\af} \metricorder \psi_{\af}(\envset')$. Finally $\psi_{\af} \in \Reach(\Rule) \subseteq \IReach(\Rule)$ by \eqref{eq:hrdp-profile}, \Cref{thm:hrdp-bounds-rdp}, \Cref{thm:solo-dominates} and \Cref{prop:ideal-profile}.
\end{proof}

Hence if $\bar\psi_{\af} \in \widehat{\Reach}_X(\Rule)$, then $\mathop{\downarrow} \psi_{\af} \subseteq \mathop{\downarrow} \bar\psi_{\af} \subseteq \widehat{\Reach}_X(\Rule)$, so the former cannot enlarge the realised profile and is skipped~(\Cref{fig:majorant-removal}). 
During the evaluation of a former, \Cref{lem:former-bounds} confines $\psi_{\af}$ to $\mathop{\downarrow} \psi_{\af}(\envset') \cap \IReach(\Rule)$, so the former can enlarge the realised profile by at most the part of this set outside the realised profile. We call the measure of this part the \emph{gap}.

\begin{definition}[Gap of a former]
    \label{def:gap}
    Let $\Sigma \subseteq \pow{\metricset}$ be a $\sigma$-algebra containing every principal lower set of $\metricposet$, and let $\mu : \Sigma \to [0, \infty]$ be a measure that is finite on principal lower sets. For $\Rule \subseteq \bset$, $X \subseteq \agentformer_{\mathrm{adm}}(\Rule)$, $\af \in \agentformer_{\mathrm{adm}}(\Rule)$ and $\envset' \subseteq \envset$, the \emph{gap} of $\af$ on $\envset'$ given $X$ is
    \[
        \Delta_{\envset'}(\af; X) \;\coloneqq\; \mu\Bigl(\bigl(\mathop{\downarrow} \psi_{\af}(\envset') \cap \IReach(\Rule)\bigr) \setminus \widehat{\Reach}_X(\Rule)\Bigr).
    \]
\end{definition}
Under \Cref{ass:finite}, $\IReach(\Rule)$, $\widehat{\Reach}_X(\Rule)$ and $\Reach(\Rule)$ are finite intersections and unions of principal lower sets, so every set measured below lies in $\Sigma$ and has finite measure. The measure $\mu$ is a design choice.
\begin{example}[Measures on $\metricset$]
    \label{ex:area-measures}
    When $\metricset$ is a product of bounded real intervals, as for the performance-cost lattice $\perfset \times \resourceset^{op}$ of \Cref{ex:perf-cost-metric} with $\perfset, \resourceset \subseteq \bbR$ bounded, the Lebesgue measure applies. When $\metricset$ is finite, as when performance and cost are graded on finite ordinal scales, the counting measure applies, and the gap is the number of performances the former can still add.
\end{example}
By \Cref{lem:former-bounds} the gap bounds the measure of what the former can still add to the realised profile, and it is non-increasing in $\envset'$ and in $X$. We therefore discard a former once its gap is below a threshold $\varepsilon \ge 0$, losing measure less than $\varepsilon$ (\Cref{fig:gap-continue,fig:gap-discard}).

$\textsc{Evaluate}$ (\Cref{alg:evaluate}) applies this logic to each admissible former. 
It skips a former if $\bar\psi_{\af} \in \widehat{\Reach}_X(\Rule)$, and otherwise evaluates it in successive environments, discarding it as soon as its gap is below $\varepsilon$. A former evaluated in every environment joins $X$. Evaluating first the formers with the largest $\mu(\mathop{\downarrow} \bar\psi_{\af} \cap \IReach(\Rule))$ is a heuristic that grows the realised profile quickly, so that more of the remaining formers are skipped or discarded early.

\begin{algorithm}[tb]
\caption{$\textsc{Evaluate}(\Rule, \varepsilon)$, the profile of a rule with early stopping}
\label{alg:evaluate}
\begin{algorithmic}[1]
\Require rule $\Rule$; admissible formers $\agentformer_{\mathrm{adm}}(\Rule)$; single-agent behaviours $\simmapk{1}(\{\agent\}, \env)$; ideal profile $\IReach(\Rule)$; threshold $\varepsilon \ge 0$
\Ensure $X \subseteq \agentformer_{\mathrm{adm}}(\Rule)$ and $\widehat{\Reach}_X(\Rule) \subseteq \Reach(\Rule)$
\State $X \gets \emptyset$
\ForAll{$\af \in \agentformer_{\mathrm{adm}}(\Rule)$}
    \State $\agent_\env \gets \af(\env, \Rule)$ for every $\env \in \envset$
    \If{$\bar\psi_{\af} \in \widehat{\Reach}_X(\Rule)$}
        \State \textbf{continue} \Comment{skip former}
    \EndIf
    \State $\envset' \gets \emptyset$
    \ForAll{$\env \in \envset$}
        \If{$\Delta_{\envset'}(\af; X) < \varepsilon$} 
            \State \textbf{break} \Comment{skip remaining environments}
        \EndIf
        \State call $\simmap(\{\agent_\env\}, \env)$ \Comment{skipped if this call was made before}
        \State $\envset' \gets \envset' \cup \{\env\}$
    \EndFor
    \If{$\envset' = \envset$} $X \gets X \cup \{\af\}$
    \EndIf
\EndFor
\State \Return $X$, $\widehat{\Reach}_X(\Rule)$
\end{algorithmic}
\end{algorithm}

$\textsc{Evaluate}$ is sound in that every performance in the realised profile is attained or dominated by a former it has evaluated in every environment, so no spurious performance enters the profile. Furthermore, the approximation error from the discarded formers is bounded, so on termination the price paid for the approximation is known.
\begin{proposition}[Soundness and approximation error of \textsc{Evaluate}]
    \label{prop:evaluate}
    Let $X$ be the set returned by \Cref{alg:evaluate} for $\Rule$ and $\varepsilon$, and let $D \subseteq \agentformer_{\mathrm{adm}}(\Rule)$ be the formers discarded for their gap.
    \begin{enumerate}
        \item \emph{Soundness.} $\widehat{\Reach}_X(\Rule) \subseteq \Reach(\Rule)$.
        \item \emph{Approximation error.} $\mu\bigl(\Reach(\Rule) \setminus \widehat{\Reach}_X(\Rule)\bigr) \le |D|\,\varepsilon$, with $\widehat{\Reach}_X(\Rule) = \Reach(\Rule)$ when $D = \emptyset$.
    \end{enumerate}
\end{proposition}
\begin{proof}
    \emph{(i)} A former enters $X$ after a call in every environment, so $\psi_{\af}$ is known for $\af \in X$, and $\widehat{\Reach}_X(\Rule) \subseteq \Reach(\Rule)$ since $\Reach(\Rule)$ is a lower set containing these performances.
    \emph{(ii)} For the bound, $\Reach(\Rule) \setminus \widehat{\Reach}_X(\Rule) \subseteq \bigcup_{\af \in \agentformer_{\mathrm{adm}}(\Rule)} \bigl(\mathop{\downarrow} \psi_{\af} \setminus \widehat{\Reach}_X(\Rule)\bigr)$, and by subadditivity it suffices to bound each term. Let $X_{\af} \subseteq X$ be the set held when $\af$ was processed. If $\af \in X$ the term is empty. If $\af$ was skipped, then $\bar\psi_{\af} \in \widehat{\Reach}_{X_{\af}}(\Rule)$ and \Cref{lem:former-bounds} gives $\mathop{\downarrow} \psi_{\af} \subseteq \mathop{\downarrow} \bar\psi_{\af} \subseteq \widehat{\Reach}_X(\Rule)$, so the term is empty. If $\af \in D$ with $\envset'$ the environments called at discard, \Cref{lem:former-bounds} gives $\mathop{\downarrow} \psi_{\af} \setminus \widehat{\Reach}_X(\Rule) \subseteq \bigl(\mathop{\downarrow} \psi_{\af}(\envset') \cap \IReach(\Rule)\bigr) \setminus \widehat{\Reach}_{X_{\af}}(\Rule)$, whose measure is $\Delta_{\envset'}(\af; X_{\af})$ $< \varepsilon$. Summing over $D$ gives the bound.
\end{proof}

With $\varepsilon = 0$ no former is discarded, so the evaluation is exact for every $\mu$. Notably, $\textsc{Evaluate}$ returns a lower set inside $\Reach(\Rule)$, which is required for the pruning of \Cref{sec:bnb}.

\subsection{\textsc{RuleBnB}}
\label{sec:rulebnb}

\textsc{RuleBnB} (\Cref{alg:rule-bnb}) receives the candidate rules, their admissible formers and a threshold $\varepsilon \ge 0$. 
It first computes the single-agent behaviours and then repeatedly selects a maximal unvisited rule. 
If the rule is pruned, it leaves the unvisited set $U$ together with its down-set, and otherwise $\textsc{Evaluate}$ returns its realised profile and the rule joins the evaluated set $V$. 
Pruning is down-closed (\Cref{lem:safe-pruning}(iii)), so pruning a maximal rule removes its whole down-set at once and may discard many candidates in a single comparison.

\textsc{RuleBnB} is an anytime algorithm. At any point of its execution, the evaluated rules give a lower bound to the attainable profile. Since the upper bound to the profile is known, we can compute the difference between the two, which we call the \emph{residual}. The residual bounds what evaluating further rules can still gain, up to the approximation error of the rules already evaluated (\Cref{cor:certificate}).
\begin{definition}[Residual]
    \label{def:residual}
    Let $V \subseteq \ruleset$ be the evaluated rules, with realised profiles $\widehat{\Reach}(\Rule)$ for $\Rule \in V$, and $U \subseteq \ruleset$ the unvisited rules. The \emph{residual} is
    \[
        G \;\coloneqq\; \bigcup_{\Rule \in \max U} \IReach(\Rule) \;\setminus\; \bigcup_{\Rule \in V} \widehat{\Reach}(\Rule).
    \]
\end{definition}

\begin{algorithm}[H]
\caption{Rule Branch and Bound (\textsc{RuleBnB})}
\label{alg:rule-bnb}
\begin{algorithmic}[1]
\Require candidate rules $\ruleset$; admissible formers $\agentformer_{\mathrm{adm}}(\Rule)$ for every $\Rule \in \ruleset$; threshold $\varepsilon \ge 0$
\Ensure evaluated rules $V$ with $\{\widehat{\Reach}(\Rule)\}_{\Rule \in V}$; unvisited rules $U$; residual $G$
\State compute $\simmapk{1}(\{\agent\}, \env)$ for every $(\agent, \env) \in \agentset \times \envset$ and $\IReach(\Rule)$ for every $\Rule \in \ruleset$ \Comment{single-agent behaviours}
\State $V \gets \emptyset$;\quad $U \gets \ruleset$;\quad $G \gets \bigcup_{\Rule \in \max \ruleset} \IReach(\Rule)$
\While{$U \neq \emptyset$}
    \State $\Rule \gets$ an element of $\max U$
    \If{$\textsc{Prune}(\Rule, V)$}
        \State $U \gets U \setminus \mathop{\downarrow}\Rule$ \Comment{sound by \Cref{lem:safe-pruning}}
    \Else
        \State $\widehat{\Reach}(\Rule) \gets \textsc{Evaluate}(\Rule, \varepsilon)$ \Comment{sound by \Cref{prop:evaluate}(i)}
        \State $V \gets V \cup \{\Rule\}$;\quad $U \gets U \setminus \{\Rule\}$
    \EndIf
    \State $G \gets \bigcup_{\Rule \in \max U} \IReach(\Rule) \setminus \bigcup_{\Rule \in V} \widehat{\Reach}(\Rule)$ \Comment{residual}
\EndWhile
\State \Return $V$, $\{\widehat{\Reach}(\Rule)\}_{\Rule \in V}$, $U$, $G$
\end{algorithmic}
\end{algorithm}

\begin{proposition}[Correctness and anytime certificate]
    \label{prop:bnb}
    \Cref{alg:rule-bnb} terminates after at most $|\ruleset|$ iterations. At any point of its execution, every feasible $\Rule \in \ruleset \setminus (V \cup U)$ satisfies $\Reach(\Rule) \subseteq \Reach(\Rule')$ for some feasible $\Rule' \in V$, and
    \begin{equation}
        \label{eq:certificate}
        \bigcup_{\Rule \in V} \widehat{\Reach}(\Rule)
        \;\subseteq\;
        \bigcup_{\Rule \in \ruleset} \Reach(\Rule),
        \qquad
        \bigcup_{\Rule \in \ruleset^*} \Reach(\Rule)
        \;\subseteq\;
        \bigcup_{\Rule \in V} \Reach(\Rule) \;\cup\; \bigcup_{\Rule \in \max U} \IReach(\Rule).
    \end{equation}
\end{proposition}
\begin{proof}
    Each iteration removes at least the chosen rule from $U$, so the loop runs at most $|\ruleset|$ times. Every $\Rule \in \ruleset \setminus (V \cup U)$ was removed as a member of $\mathop{\downarrow}\Rule_1$ for some $\Rule_1$ with $\textsc{Prune}(\Rule_1, V_t)$ true at that time, where $V_t \subseteq V$ is the set then evaluated, and since $\widehat{\Reach}$ is fixed at evaluation, \Cref{lem:safe-pruning}(iii) gives $\textsc{Prune}(\Rule, V_t)$. If $\Rule$ is feasible, \Cref{lem:safe-pruning}(ii) rules out the first clause of \eqref{eq:prune}, so \Cref{lem:safe-pruning}(i) gives $\Reach(\Rule) \subseteq \Reach(\Rule')$ for some $\Rule' \in V_t \subseteq V$, and $\Rule'$ is feasible since $L^* \subseteq \Reach(\Rule) \subseteq \Reach(\Rule')$.

    The first inclusion of \eqref{eq:certificate} holds since $\widehat{\Reach}(\Rule) \subseteq \Reach(\Rule)$. For the second let $\Rule \in \ruleset^*$. If $\Rule \in V$ there is nothing to show. If $\Rule \in U$, pick $\Rule_0 \in \max U$ with $\Rule \subseteq \Rule_0$, so that $\Reach(\Rule) \subseteq \IReach(\Rule) \subseteq \IReach(\Rule_0)$ by \Cref{thm:solo-dominates,prop:ideal-profile}. Otherwise the first paragraph gives $\Reach(\Rule) \subseteq \Reach(\Rule')$ for some $\Rule' \in V$.
\end{proof}

\begin{corollary}[Computable certificate]
    \label{cor:certificate}
    At any point of the execution of \Cref{alg:rule-bnb}, with $D_{\Rule}$ the formers discarded by $\textsc{Evaluate}(\Rule, \varepsilon)$ for $\Rule \in V$,
    \[
        \mu\Bigl(\, \bigcup_{\Rule \in \ruleset^*} \Reach(\Rule) \;\setminus\; \Bigl( \bigcup_{\Rule \in V} \widehat{\Reach}(\Rule) \cup G \Bigr) \Bigr) \;\le\; \varepsilon \sum_{\Rule \in V} |D_{\Rule}|.
    \]
\end{corollary}
\begin{proof}
    By \eqref{eq:certificate} the set on the left is contained in $\bigcup_{\Rule \in V} \bigl(\Reach(\Rule) \setminus \widehat{\Reach}(\Rule)\bigr)$, and \Cref{prop:evaluate}(ii) bounds the measure of each term by $|D_{\Rule}|\,\varepsilon$.
\end{proof}

At termination $U = \emptyset$. With $\varepsilon = 0$ the realised profiles are exact (\Cref{prop:evaluate}(ii)), and by \Cref{prop:bnb} every feasible candidate is dominated by a feasible evaluated rule. The maximal profiles among the feasible evaluated rules are therefore the maximal profiles among all feasible candidates, so \textsc{RuleBnB} solves \eqref{prob:BRDP} over $\ruleset$, and \eqref{prob:RDP} with $L^* = \{\bot_{\metricset}\}$. Since pruning may discard a rule whose profile equals that of an evaluated one, $V$ contains every solution profile but not necessarily every solution.

For $\varepsilon > 0$ the realised profiles are subsets of the true ones, so a rule reported feasible is feasible, whereas a feasible rule may be reported infeasible, missing $L^*$ by a set of measure at most $|D_{\Rule}|\,\varepsilon$ (\Cref{prop:evaluate}(ii)). Every feasible candidate is still dominated by a feasible evaluated rule $\Rule'$ (\Cref{prop:bnb}), whose realised profile falls short of it by at most $|D_{\Rule'}|\,\varepsilon$ in measure, but which candidates are maximal is not certified.

 \section{Case Study: Designing Rules for Robotaxis}
\label{sec:case-studies}

We now apply \textsc{RuleBnB} to solve \eqref{prob:HRDP} for the robotaxi problem of~\Cref{sec:rdp_inf_ex2}.
The case study serves three purposes.
First, it demonstrates how to pose and solve the~\gls{rdp} on a problem that resembles a typical regulatory task.
Second, it illustrates \Cref{thm:solo-dominates,thm:optimal-monotone} and exhibits an instance of \Cref{thm:nonoptimal}, where the catalogue is incomplete and the most permissive rule is not optimal.
Third, it shows that the solution is interpretable, for the regulator and for the designer alike: the optimal rule tells the regulator which restrictions must be imposed for the agents to operate safely and complete the tasks required by the scenario, and the behaviours it excludes tell the designer what limits system performance, that is, what the agents find challenging.

\subsection{Setup}
\begin{figure}[H]
    \centering
    \begin{subfigure}[b]{0.2\textwidth}
        \centering
        \includegraphics[width=\linewidth]{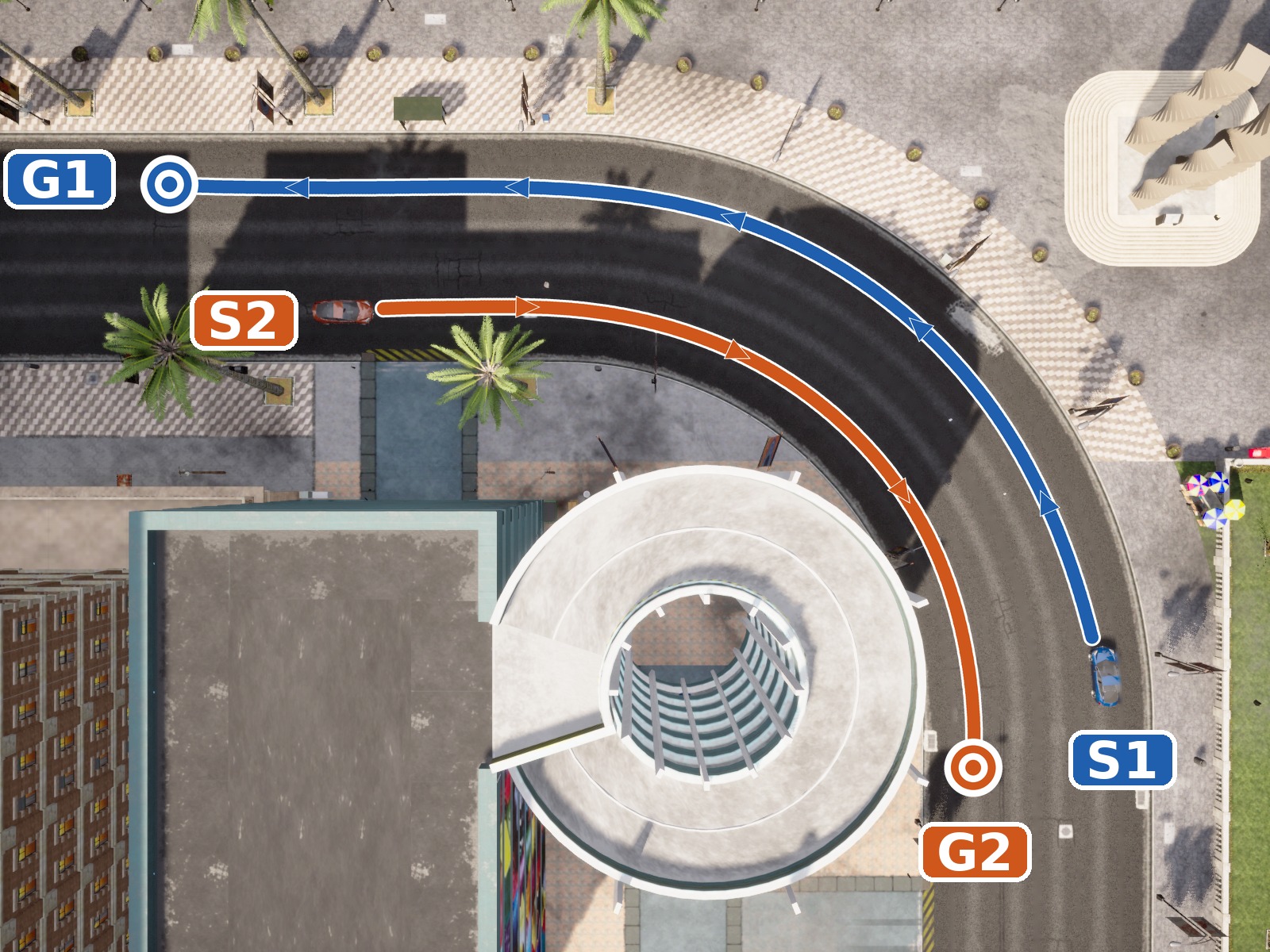}
        \caption{Curve.}
        \label{fig:av-curve}
    \end{subfigure}
    \hspace{0.04\textwidth}
    \begin{subfigure}[b]{0.2\textwidth}
        \centering
        \includegraphics[width=\linewidth]{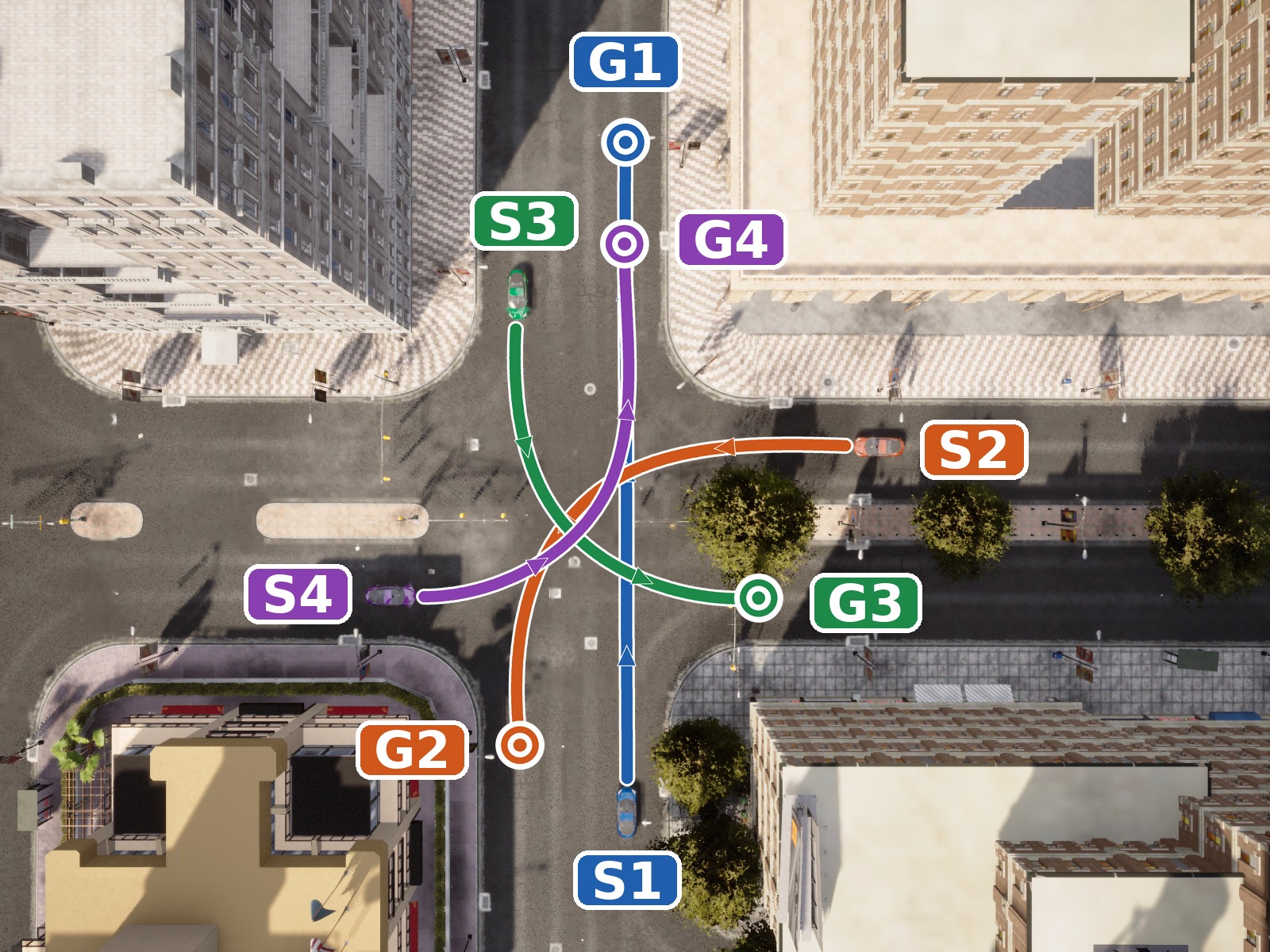}
        \caption{Intersection.}
        \label{fig:av-intersection}
    \end{subfigure}
    \caption{The two scenarios. Vehicle $i$ starts at $S_i$ and ends at $G_i$. The drawn line is a possible trajectory; the trajectory executed depends on the agent and on the rule in force.}
    \label{fig:av-scenarios}
\end{figure}

\paragraph{Environments.}
The scenes are a curve and an intersection, denoted by $\scene_{\mathrm{C}},\scene_{\mathrm{J}}$ respectively, with capacities of two and four vehicles respectively (\Cref{fig:av-scenarios}). They give two instances of \eqref{prob:HRDP}, with environment sets
\[
    \envset_{\mathrm{C}} \coloneqq \{(\scene_{\mathrm{C}}, n)\}_{n=1}^{2}, \qquad \envset_{\mathrm{J}} \coloneqq \{(\scene_{\mathrm{J}}, n)\}_{n=1}^{4}.
\]

\paragraph{Performance.}
Each scene assigns a fixed task to vehicle $i$, from $S_i$ to $G_i$ (\Cref{fig:av-scenarios}), so a joint behaviour of $j$ vehicles carries the tasks of vehicles $1, \dots, j$. The performance takes the form of \Cref{ex:perf-cost-metric}, a pair of travel-time efficiency and cost,
\[
    \perf(\agent, \beh, \env) = \Bigl( \min_{i \in [j]} \frac{t^*_i(\env)}{t_i(\beh, \env)},\; \Cost(\agent) \Bigr),
\]
where $t_i(\beh, \env)$ is the travel time of vehicle $i$ under the joint behaviour $\beh$ and $t^*_i(\env)$ is the optimal travel time of its task, taken as the length of the shortest path driven at the maximum speed of the vehicles, $60\,\mathrm{km/h}$. Here $j \le n$ is the number of vehicles in the joint behaviour $\beh$, and $t_i(\beh, \env) = \infty$, hence efficiency $0$, when vehicle $i$ does not reach $G_i$. The operator has a budget for purchasing and operating the vehicles, so $\Cost(\agent)$ is the price of the components of the agent. 

\paragraph{Agents, Formers and the Catalogue.}
An agent is composed of perception, actuation and policy subsystems. Actuation is fixed throughout and relies on the physics engine of CARLA~\cite{pmlr-v78-dosovitskiy17a}. We vary perception and policy. 
The perception system is either a single lidar with range $30$ or $80\,\mathrm{m}$ and a field of view of $90$ or $360$ degrees, or a pair of lidars, one at the front and one at the rear, each with range $30$ or $80\,\mathrm{m}$ and a field of view of $180$ degrees. 
The policy is a minimum-time nonlinear model predictive controller~\cite{Gros2020RealTimeIteration}, solved online by using sequential quadratic programming, which predicts the occupancy of the other vehicles by their reachable set intersected with the rule or, with the two-lidar perceptions, by their shadow occupancy intersected with the rule\footnote{The shadow occupancy accounts for blind spots by inserting shadow vehicles in those regions~\cite{Nager2019}.}. 
This gives six agents for each rule.
A deployed agent has fixed parameters, that is, the set of agents is given by the above combination of subsystems initialised for every rule considered.
An agent former selects which agent to deploy in each environment and under each rule, and may in particular deploy an agent initialised for a rule other than the one in force. We use two catalogues. The \emph{complete} catalogue consists of every function from environments and rules to agents, and is complete in the sense of \Cref{def:complete-catalogue-general}, since for every admissible deployment it contains a former that returns one agent of each pool, which is admissible and performs at least as well (\Cref{lem:perf-pool}). The \emph{incomplete} catalogue contains one former per combination of subsystems, which in every environment deploys that combination initialised for the rule in force. Such a former neither varies with the environment nor deploys an agent initialised for a different rule, and the catalogue is therefore not complete.

\begin{figure}[H]
    \centering
    \begin{subfigure}[b]{0.3\textwidth}
        \centering
        \includegraphics[width=\linewidth]{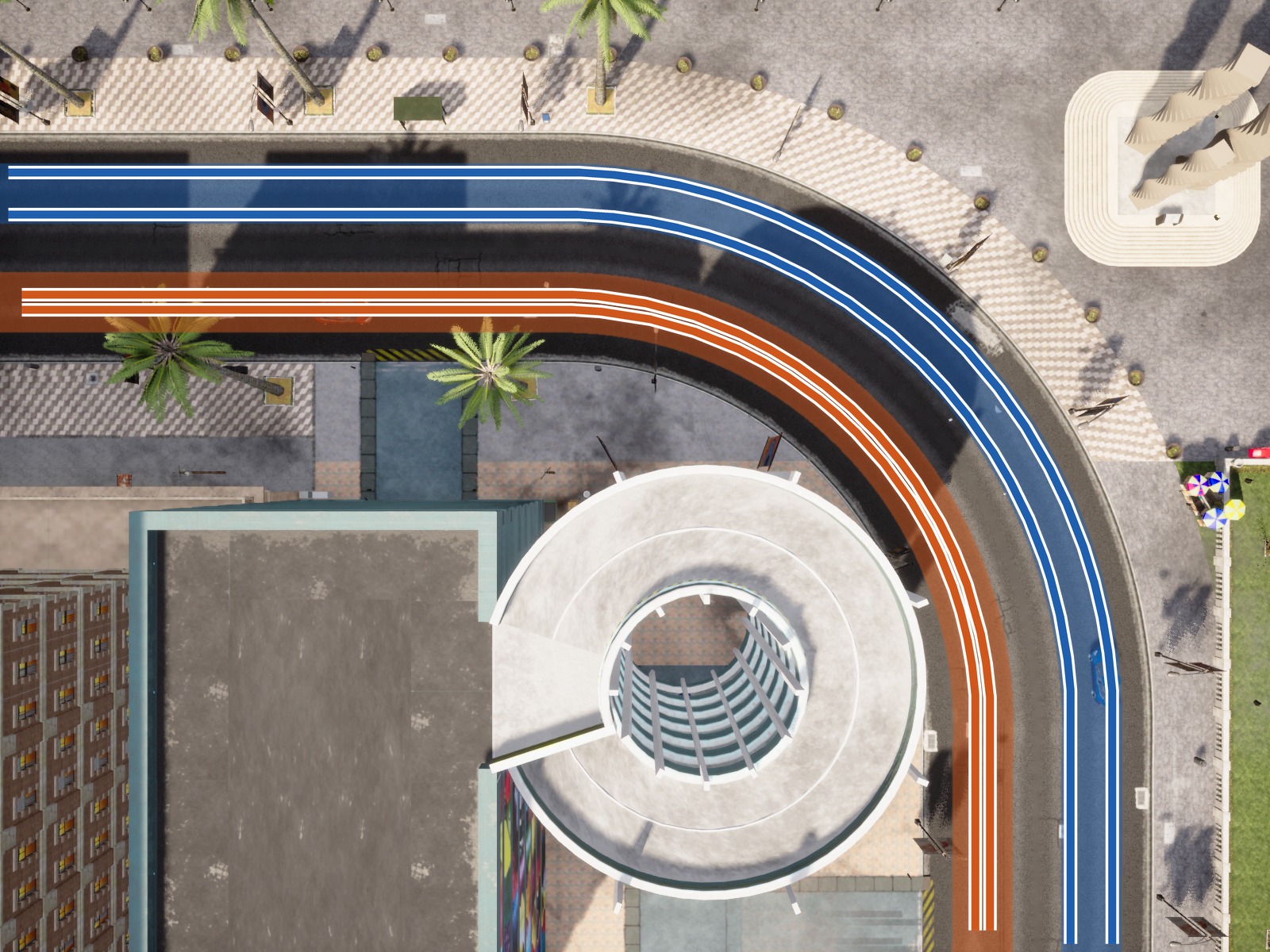}
        \caption{$c = 5\,\mathrm{m}$.}
        \label{fig:av-carriageway-c5}
    \end{subfigure}
    \hspace{0.02\textwidth}
    \begin{subfigure}[b]{0.3\textwidth}
        \centering
        \includegraphics[width=\linewidth]{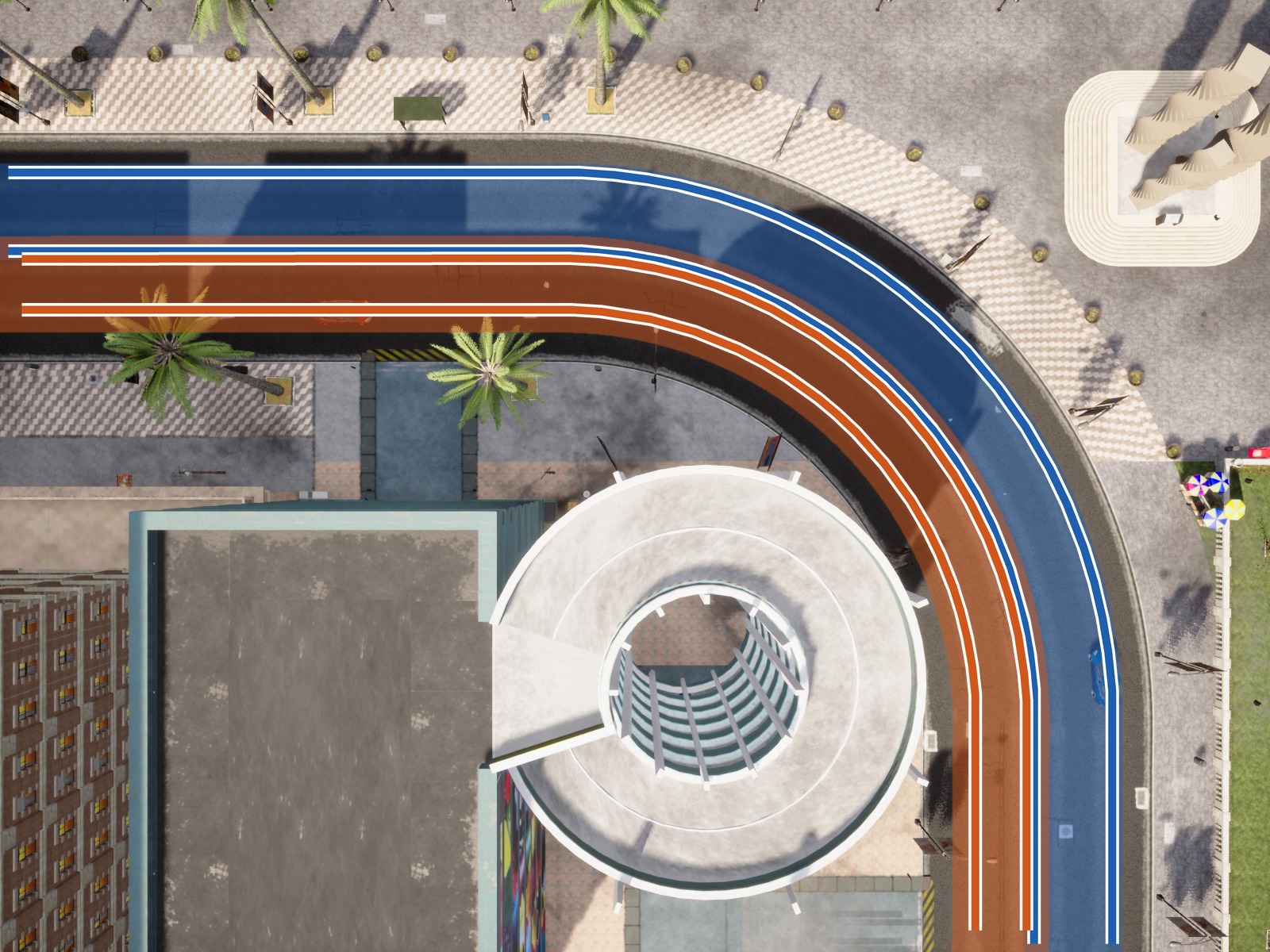}
        \caption{$c = 8\,\mathrm{m}$.}
        \label{fig:av-carriageway-c8}
    \end{subfigure}
    \hspace{0.02\textwidth}
    \begin{subfigure}[b]{0.3\textwidth}
        \centering
        \includegraphics[width=\linewidth]{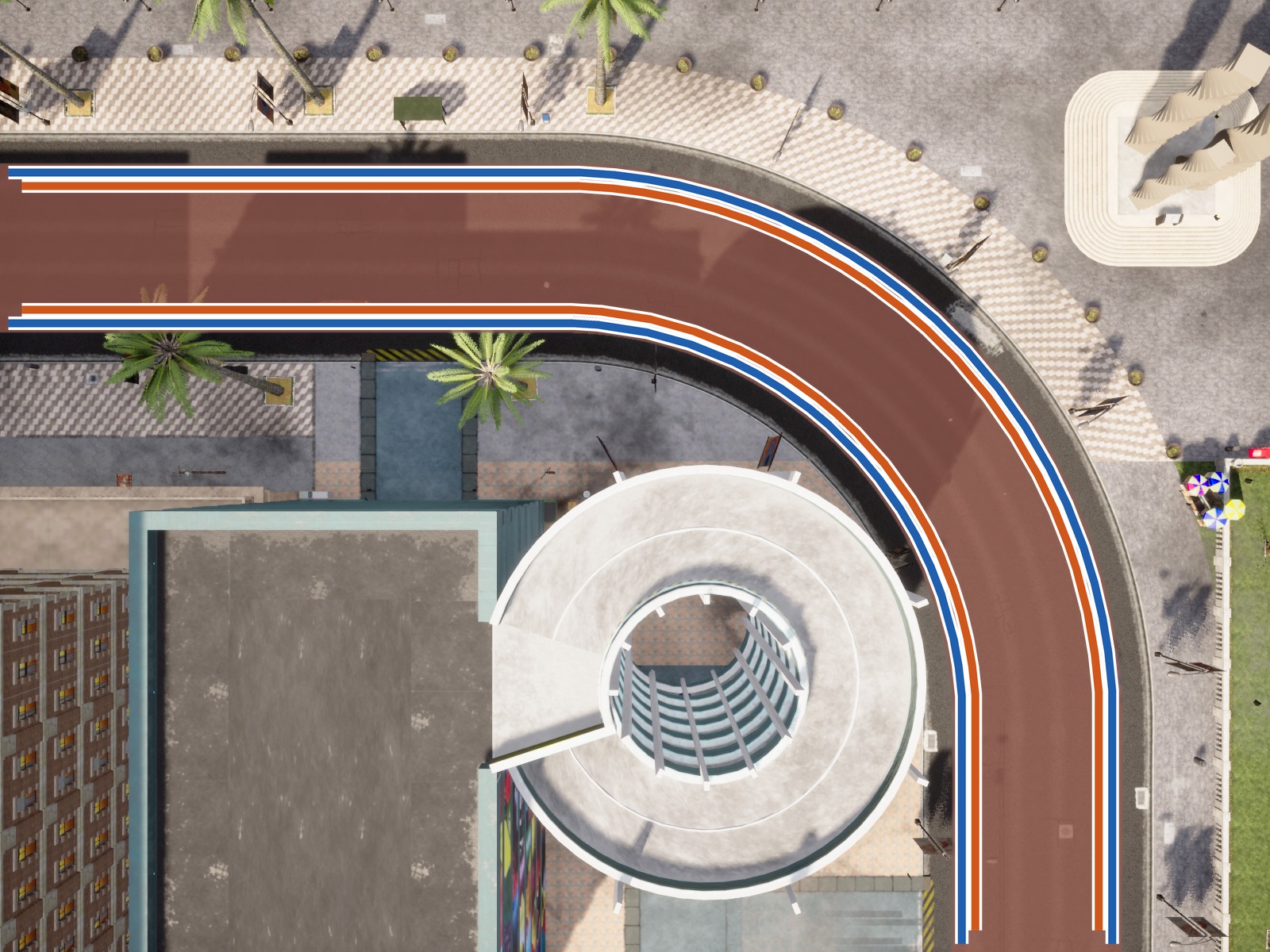}
        \caption{$c = 14\,\mathrm{m}$.}
        \label{fig:av-carriageway-c14}
    \end{subfigure}
    \caption{The carriageway rule on the curve. Each vehicle may occupy the region shaded and outlined in its colour, driving in its direction of travel; the blend is the region both may occupy, which appears from $c = 8\,\mathrm{m}$.}
    \label{fig:av-carriageway}
\end{figure}

\paragraph{Rules.}
A rule is a speed limit together with a carriageway width (\Cref{fig:av-carriageway}). The candidate rules are
\[
    \ruleset = \{5, 6, 8, 10, 12, 14\}\,\mathrm{m} \times \{20, 40, 60\}\,\mathrm{km/h},
\]
ordered by inclusion of the behaviours they permit, so that a wider carriageway and a higher speed limit give a more permissive rule.

\paragraph{Solver.}
We solve \eqref{prob:HRDP} with \textsc{RuleBnB} (\Cref{alg:rule-bnb}). Following \Cref{alg:rule-bnb}, we begin with the most permissive rule, $\top = (14\,\mathrm{m}, 60\,\mathrm{km/h})$, and move down the chains of $\ruleset$. When $\max U$ is not a singleton, the choice is arbitrary and does not affect correctness; since we anticipate that the speed limit has the greater influence on the travel-time efficiency, we take the rule maximising $c + v$, and break remaining ties by the higher speed limit.

\subsection{Single-Agent Deployments Weakly Dominate Multi-Agent Ones}
\label{sec:av-capacity}
\begin{figure}[H]
    \centering
    \includegraphics{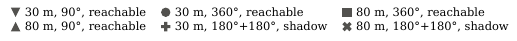}\\[0.3ex]
    \begin{subfigure}[t]{2.04in}
        \centering
        \includegraphics{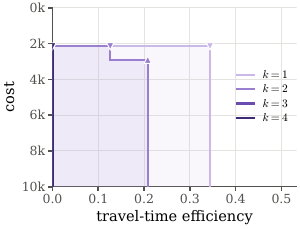}
        \caption{Profile for $k = 1, 2, 3, 4$.}
        \label{fig:av-capacity-profile}
    \end{subfigure}\hspace{1em}\begin{subfigure}[t]{1.58in}
        \centering
        \includegraphics{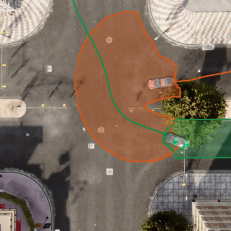}
        \caption{Two vehicles. The reachable sets do not block one another.}
        \label{fig:av-intersection-k2}
    \end{subfigure}\hspace{1em}\begin{subfigure}[t]{1.58in}
        \centering
        \includegraphics{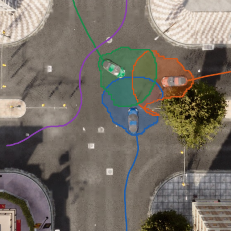}
        \caption{Four vehicles. The reachable sets block one another and the vehicles deadlock.}
        \label{fig:av-intersection-k4}
    \end{subfigure}\caption{(\subref{fig:av-capacity-profile}) The profile $\Reachk{k}(\Rule)$ of $\Rule = (14\,\mathrm{m}, 40\,\mathrm{km/h})$ on the intersection scene $\scene_{\mathrm{J}}$ for capacities $k = 1, 2, 3, 4$, under either catalogue. For $k = 3, 4$ every former attains efficiency $0$, so the profile lies on the cost axis. (\subref{fig:av-intersection-k2}, \subref{fig:av-intersection-k4}) Snapshots of the intersection scene $\scene_{\mathrm{J}}$ under the reachable-set prediction, with $k = 2$ and $k = 4$ vehicles.}
    \label{fig:av-capacity}
\end{figure}

\Cref{fig:av-capacity-profile} illustrates~\Cref{thm:solo-dominates}. It plots the profile of a fixed rule on the intersection scene as the capacity grows. Both the complete and incomplete catalogues produce the same profiles for this rule. The complete catalogue may deploy an agent initialised for a stricter rule, but here no such agent improves the formers on the profile, the $30$ and $80\,\mathrm{m}$, $90^\circ$ reachable-set formers; it improves only the shadow formers with one vehicle, which remain dominated by cheaper and more efficient formers. The profile does not grow with the capacity, and in this case it shrinks. 

In particular, with three and four vehicles every former attains zero travel time efficiency. That is, in some environment at least one vehicle does not complete its task. This is because the reachable-set prediction is conservative, and once several vehicles meet in the intersection their predicted occupancies cover one another's paths and the vehicles deadlock (\Cref{fig:av-intersection-k2,fig:av-intersection-k4}). In fact, this is very informative from a design perspective. If the regulator or operator wishes to have greater performance, either they need to improve the capabilities of the agents to deal with behaviours in which agents interact, or they need to impose additional rules. Indeed, the rules of the road include right-of-way rules for junctions and intersections.

\subsection{The Profile is Monotone under Complete Catalogues}
\begin{figure}[H]
    \centering
    \includegraphics{img/RDP/generated/av_formers_legend.pdf}\\[0.3ex]
    \begin{subfigure}[b]{1.65in}
        \centering
        \includegraphics{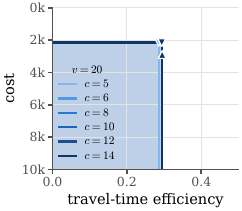}
        \caption{$v = 20\,\mathrm{km/h}$, varying $c$.}
        \label{fig:av-monotone-c}
    \end{subfigure}\hspace{0.6em}\begin{subfigure}[b]{1.65in}
        \centering
        \includegraphics{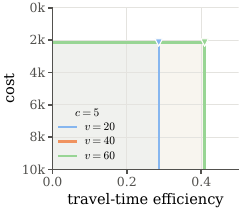}
        \caption{$c = 5\,\mathrm{m}$, varying $v$.}
        \label{fig:av-monotone-v}
    \end{subfigure}\hspace{0.6em}\begin{subfigure}[b]{2.41in}
        \centering
        \includegraphics{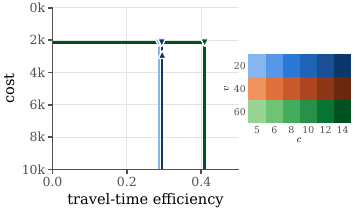}
        \caption{All rules.}
        \label{fig:av-monotone-all}
    \end{subfigure}\caption{Profiles on the curve, $\envset_{\mathrm{C}}$, under the complete catalogue. Larger profiles extend further to the right (more efficient) and upwards (cheaper). Profiles that coincide are drawn on top of one another: for $v = 20\,\mathrm{km/h}$ every $c \ge 6\,\mathrm{m}$ attains the same profile, and for $v \in \{40, 60\}\,\mathrm{km/h}$ every rule does. The hue of a profile gives its speed limit and its shade the carriageway width, darker for wider; the key in (\subref{fig:av-monotone-all}) shows all rules. The markers are the formers attaining each corner of a profile (legend above).}
    \label{fig:av-monotone}
\end{figure}

\Cref{thm:optimal-monotone} states that the profile is monotone under a complete catalogue. \Cref{fig:av-monotone} plots the profile for varying rules under a complete catalogue. The profile is monotone, as the theorem requires. At $v = 20\,\mathrm{km/h}$ the profile grows from $c = 5$ to $6\,\mathrm{m}$ and is constant beyond. Every rule with $v \in \{40, 60\}\,\mathrm{km/h}$ attains the same profile, which contains every profile at $v = 20\,\mathrm{km/h}$, so these rules, $\top$ among them, are the maximal rules.

\begin{figure}[H]
    \centering
    \begin{subfigure}[b]{1.62in}
        \centering
        \includegraphics{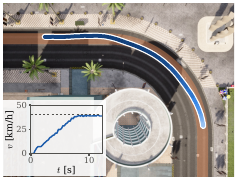}
        \caption{$c = 5\,\mathrm{m}$ ($12.45\,\mathrm{s}$).}
        \label{fig:av-traj-c5}
    \end{subfigure}\hspace{1.2em}\begin{subfigure}[b]{1.62in}
        \centering
        \includegraphics{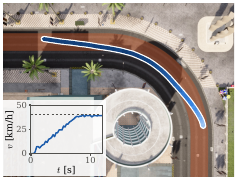}
        \caption{$c = 8\,\mathrm{m}$ ($12.15\,\mathrm{s}$).}
        \label{fig:av-traj-c8}
    \end{subfigure}\hspace{1.2em}\begin{subfigure}[b]{1.62in}
        \centering
        \includegraphics{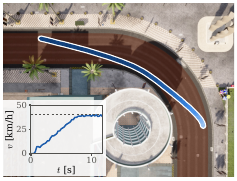}
        \caption{$c = 14\,\mathrm{m}$ ($11.80\,\mathrm{s}$).}
        \label{fig:av-traj-c14}
    \end{subfigure}\caption{A single vehicle on the outer lane of the curve under $v = 40\,\mathrm{km/h}$ (former: $80\,\mathrm{m}$, $360^\circ$ lidar, reachable-set prediction), coloured by speed, with its speed profile (inset) and, in brackets, its arrival time. The shaded region is where the carriageway rule permits the vehicle, driving in its direction of travel, to be. A wider carriageway lets the vehicle cut the corner and arrive sooner. This is a single-vehicle effect: the profile at $v = 40\,\mathrm{km/h}$ is the same for every $c$ (\Cref{fig:av-monotone}), since the worst case over $\envset_{\mathrm{C}}$ is attained with two vehicles.}
    \label{fig:av-trajectories}
\end{figure}

When keeping the speed limit fixed, a larger carriageway results in a larger profile. This is because it permits vehicles to cut the corner and arrive sooner (\Cref{fig:av-trajectories}).

However, every rule with $v \in \{40, 60\}\,\mathrm{km/h}$ attains the same profile. This is because the speed limit has the greatest influence over the travel time efficiency of the vehicles, but for such short routes the vehicles are not able to accelerate past $40\,\mathrm{km/h}$. 

A non-unique solution to \eqref{prob:HRDP} has an interesting implication from a regulatory perspective. Since every rule with $v \in \{40, 60\}\,\mathrm{km/h}$ attains the same profile, a regulator could pick any of them. However, recall that part of the goal of designing rules was to maximise the performance of future agents. Intuitively, the closure of the system under a more permissive rule may admit better-performing agents in the future, which would favour the $60\,\mathrm{km/h}$ limit; whether it does can only be settled once the closure can be computed.

\subsection{The Profile is not Monotone under Incomplete Catalogues}
Under a complete catalogue and worst-case aggregation, the most permissive rule $\Rule = \top$ solves \eqref{prob:HRDP}. In practice catalogues are rarely complete. An incomplete catalogue thus corresponds to formers that cannot always deploy the best agent available for the environment and the rule.

We now show that when the catalogue is not complete, the rule attaining a maximal profile need not be the most permissive.

\begin{figure}[H]
    \centering
    \includegraphics{img/RDP/generated/av_formers_legend.pdf}\\[0.3ex]
    \begin{minipage}[b]{2.49in}
        \begin{subfigure}[b]{\linewidth}
            \centering
            \includegraphics{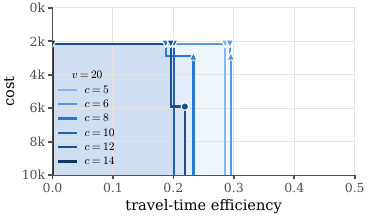}
            \caption{$v = 20\,\mathrm{km/h}$.}
            \label{fig:av-nonmono-v20}
        \end{subfigure}\\[1ex]
        \begin{subfigure}[b]{\linewidth}
            \centering
            \includegraphics{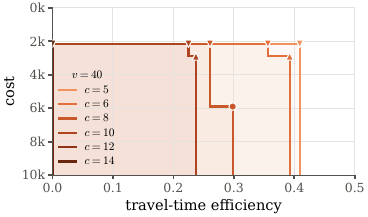}
            \caption{$v = 40\,\mathrm{km/h}$.}
            \label{fig:av-nonmono-v40}
        \end{subfigure}\end{minipage}\hspace{2em}\begin{subfigure}[b]{2.17in}
        \centering
        \includegraphics{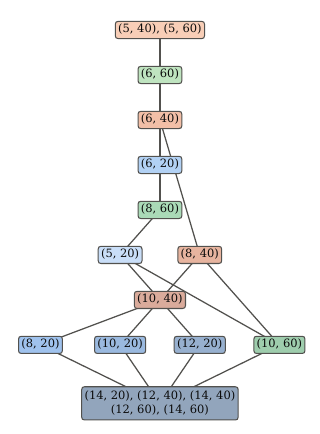}
        \caption{All rules.}
        \label{fig:av-nonmono-hasse}
    \end{subfigure}\caption{Profiles on the curve, $\envset_{\mathrm{C}}$, under the incomplete catalogue. (\subref{fig:av-nonmono-v20}, \subref{fig:av-nonmono-v40}) The profile of each carriageway width $c$ for a fixed speed limit. (\subref{fig:av-nonmono-hasse}) The Hasse diagram of $\ruleset$ ordered by profile; rules with equal profiles share a node. Rules are written $(c, v)$.}
    \label{fig:av-nonmonotone}
\end{figure}

\Cref{fig:av-nonmonotone} shows the profiles under the incomplete catalogue. They are not monotone in $c$. For example, at $v = 40$ and $60\,\mathrm{km/h}$ the profiles form a chain whose profile shrinks as $c$ grows, reaching the smallest profile with $c = 12\,\mathrm{m}$.

This occurs because of how the predictions module works. Each former predicts the other vehicle by its reachable set intersected with the rule in force. The wider the carriageway, the more of the road the rule permits the oncoming vehicle to occupy (\Cref{fig:av-carriageway}), and so the occupancy each vehicle predicts for the other covers more of its own path.

\subsection{Solving the HRDP with \textsc{RuleBnB}}
\label{sec:av-solving}
\begin{figure}[H]
    \centering
    \includegraphics{img/RDP/generated/av_formers_legend.pdf}\\[0.3ex]
    \begin{subfigure}[b]{2.32in}
        \centering
        \includegraphics{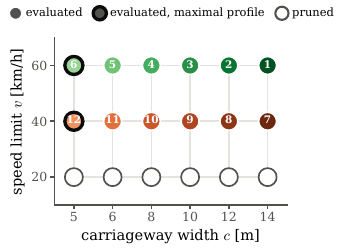}
        \caption{Evaluated and pruned rules.}
        \label{fig:av-bnb-hasse}
    \end{subfigure}\hspace{1.2em}\begin{subfigure}[b]{1.79in}
        \centering
        \includegraphics{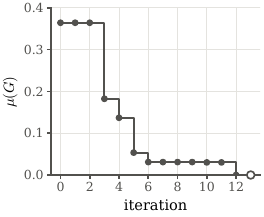}
        \caption{Residual.}
        \label{fig:av-bnb-residual}
    \end{subfigure}\hspace{1.2em}\begin{subfigure}[b]{1.87in}
        \centering
        \includegraphics{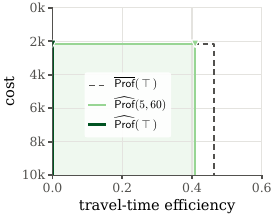}
        \caption{Realised profiles.}
        \label{fig:av-bnb-profiles}
    \end{subfigure}\\[1.5ex]
    \begin{subfigure}[b]{4.97in}
        \centering
        \includegraphics{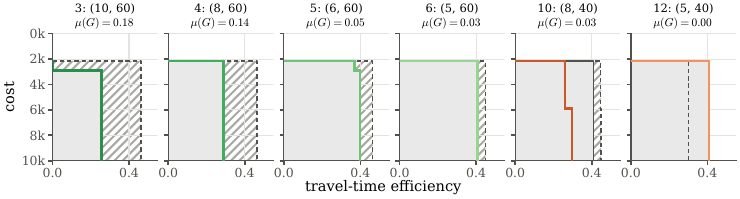}
        \caption{Construction of the realised profiles.}
        \label{fig:av-bnb-progress}
    \end{subfigure}\caption{\textsc{RuleBnB} on the curve under the incomplete catalogue. (\subref{fig:av-bnb-hasse}) The rules, numbered in the order of evaluation. (\subref{fig:av-bnb-residual}) The normalised area of the residual after each iteration; the open marker is the pruning step. (\subref{fig:av-bnb-profiles}) The profiles of $(5\,\mathrm{m}, 60\,\mathrm{km/h})$ and $\top$, with the single-agent ideal profile $\IReachk{1}(\top)$ (dashed). (\subref{fig:av-bnb-progress}) The iterations at which the residual shrinks: the profiles evaluated so far (grey), the profile just evaluated (in its colour), the ideal profiles of the unvisited rules (dashed) and the residual (hatched).}
    \label{fig:av-bnb}
\end{figure}

We now show briefly how \textsc{RuleBnB} works, on the incomplete catalogue with $\varepsilon = 0$ (\Cref{fig:av-bnb}). The algorithm starts at the most permissive rule $\top$ and descends the order iteratively (\Cref{fig:av-bnb-hasse}). It first descends the chain of $60\,\mathrm{km/h}$ rules from the widest carriageway to the narrowest, then the chain of $40\,\mathrm{km/h}$ rules. When it reaches the $20\,\mathrm{km/h}$ rules, the ideal profile of the widest of them is already contained in the profile of $(6\,\mathrm{m}, 60\,\mathrm{km/h})$. The whole $20\,\mathrm{km/h}$ chain lies below that rule and is completely pruned in one go (\Cref{lem:safe-pruning}). Of the $18$ rules, $12$ are evaluated and $6$ are pruned.

\Cref{fig:av-bnb-residual,fig:av-bnb-progress} plots the residual of the profiles evaluated at each iteration. It shrinks as the evaluated profiles cover more of the ideal profiles of the unvisited rules, and it is empty once they cover all of them.

\subsection{Implications of the RDP for the Regulator and the Agent Designer}
\label{sec:av-interpretation}
The solution to \eqref{prob:HRDP} is informative to the regulator and to the agent designer. 

For the regulator, $\Rule^* = (5\,\mathrm{m}, 60\,\mathrm{km/h})$ is a rule that maximises the profile. Although the agents always remain safe, more permissive rules reduce the travel-time efficiency, and more restrictive ones do not improve it.
Now suppose the regulator selects $\Rule^*$, and the current former $\af$, the $30\,\mathrm{m}$, $90^\circ$ reachable-set former, is running on $\envset_{\mathrm{C}}$. Once the closure $\Li_{\Rule^*}(\{\af\})$ can be computed (\Cref{sec:licensing}), any former in it may be admitted alongside $\af$ without affecting admissibility with the deployed formers.

For the designer, the solution reveals which behaviours the agents cannot yet perform. $\Rule^*$ excludes the wide carriageways, on which the reachable-set prediction is so conservative that two oncoming vehicles slow one another; a less conservative prediction could make a wider carriageway preferable.
If the designers cannot remove this conservatism, they may instead ask the regulator to consider a wider range of rules, such as right-of-way rules.

In this sense the~\gls{rdp} is a dual of the agent design problem. When no rule yields a non-empty profile, either the requirements of the operator are too strict or the agents must be improved, whereas when a rule does, the behaviours it excludes indicate to the designer which capabilities require improvement. \section{Conclusions \& Future Work}
\label{sec:future-work}
\paragraph{Conclusions} In this work we introduced the~\gls{rdp}, a system- and agent-agnostic framework for the design of rules in multi-agent systems. The~\gls{rdp} is framed from the regulatory perspective, in which, given a catalogue of agents and a system performance to maximise, one seeks the rules under which the attainable system performance is largest. We first introduced \eqref{prob:RDP} in its full generality, and then \eqref{prob:HRDP} as its homogeneous instance. By studying~\eqref{prob:HRDP} we found that, under worst-case aggregation, increasing the number of agents that may operate at once never enlarges the profile, that relaxing the rule never shrinks it when the catalogue is complete, so that the most permissive rule is then optimal, and that for an incomplete catalogue a stricter rule may strictly enlarge the profile, so that the optimal rule depends on the agents at hand. We further showed that under worst-case aggregation \eqref{prob:HRDP} and \eqref{prob:RDP} have the same solutions, permitting us to solve \eqref{prob:RDP} by formulating and solving \eqref{prob:HRDP}.

To solve the~\gls{rdp} under worst-case aggregation we introduced \textsc{RuleBnB}, an anytime branch-and-bound solver that operates over a finite set of candidate rules. \textsc{RuleBnB} bounds every candidate from above by its ideal profile, prunes the candidates that the bound shows to be dominated, evaluates the rest with a bounded approximation error, and certifies at any point of its execution how much of the attainable performance remains unexplored.

Applied to the regulation of robotaxis on a campus, we formulated a~\gls{hrdp} and used \textsc{RuleBnB} to solve it. We found that neither the most permissive nor the strictest candidate is optimal, and that the rule returned admits a direct interpretation for the regulator, as the rule to impose, and for the supplier, as the vehicles that may be deployed under it.

\paragraph{Future work}

\emph{Heterogeneous fleets and other aggregations.} In this work we studied worst-case aggregation. Some systems, however, can exploit the coupling between coordinating agents to improve system performance, as cyclists do by drafting and vehicles by platooning on highways. Such gains are invisible to the worst case and fall under other aggregations. Furthermore, multi-agent systems are often composed of heterogeneous fleets. This raises the main question for future work, namely to formulate, study and solve the \gls{rdp} under other aggregations and for heterogeneous fleets.

\emph{Beyond rules as sets.} In this work a rule is a set of permitted behaviours, and an agent either complies with it or does not. Reality is less clear-cut. Agents do not always satisfy every rule in force, and there are scenarios in which no behaviour satisfies all of them at once. How to account for violations, whether occasional or unavoidable, and how to integrate them into the \gls{rdp} remain open questions.

\bibliographystyle{IEEEtran}

\end{document}